\documentclass{article}

\usepackage{arxiv}

\usepackage[utf8]{inputenc} 
\usepackage[T1]{fontenc}    
\usepackage{hyperref}       
\hypersetup{
	colorlinks   = true, 
	urlcolor     = blue, 
	linkcolor    = blue, 
	citecolor   = blue 
}
\usepackage{url}            
\usepackage{booktabs}       
\usepackage{amsfonts}       
\usepackage{nicefrac}       
\usepackage{microtype}      
\usepackage{lipsum}
\usepackage{graphicx}
\usepackage{amsmath}
\usepackage{amsfonts}
\usepackage{amssymb}
\usepackage{multirow}
\usepackage{float}
\usepackage{adjustbox}
\usepackage{booktabs}
\usepackage{lscape}
\usepackage{enumitem, array}
\usepackage{xcolor}
\usepackage[round,authoryear]{natbib}
\usepackage{algorithm2e}
\usepackage{subcaption}
\SetAlgorithmName{Procedure}{procedure}{List of Procedures}

\usepackage{tikz}
\usepackage{bm}
\usepackage{relsize}

\usetikzlibrary{positioning}

\title{Implementation of the Critical Wave Groups Method with Computational Fluid Dynamics and \\ Neural Networks}

\author{
	Kevin M. Silva$^{1,2,\star}$ and Kevin J. Maki$^{2}$ \\
	\\
	$^1$Naval Surface Warfare Center Carderock Division, USA\\
	$^2$Department of Naval Architecture and Marine Engineering, The University of Michigan, USA\\
	$^\star$Corresponding author: \texttt{kevin.m.silva14.civ@us.navy.mil} \\
}

\begin{document}
	\maketitle
	
	\begin{abstract} Accurate and efficient prediction of extreme ship responses continues to be a challenging problem in ship hydrodynamics. Probabilistic frameworks in conjunction with computationally efficient numerical hydrodynamic tools have been developed that allow researchers and designers to better understand extremes. However, the ability of these hydrodynamic tools to represent the physics quantitatively during extreme events is limited. Previous research successfully implemented the critical wave groups (CWG) probabilistic method with computational fluid dynamics (CFD). Although the CWG method allows for less simulation time than a Monte Carlo approach, the large quantity of simulations required is cost prohibitive. The objective of the present paper is to reduce the computational cost of implementing CWG with CFD, through the construction of long short-term memory (LSTM) neural networks. After training the models with a limited quantity of simulations, the models can provide a larger quantity of predictions to calculate the probability. The new framework is demonstrated with a 2-D midship section of the Office of Naval Research Tumblehome (ONRT) hull in Sea State 7 and beam seas at zero speed. The new framework is able to produce predictions that are representative of a purely CFD-driven CWG framework, with two orders of magnitude of computational cost savings. 
	\end{abstract}

	
	\section*{Introduction}
	
	Ensuring the safety of a vessel in extreme ocean conditions is a crucial consideration for designers and operators. Designers optimize the design for normal operating conditions while ensuring that it will withstand the most extreme conditions. Due to the stochastic nature of the waves and the rarity of extreme events, identifying wave sequences that lead to extremes with a Monte-Carlo approach is expensive. Different probabilistic frameworks have been developed both to identify extremes and calculate the probability of their occurrence. These probabilistic methods include extrapolation-type approaches such as Peaks-Over-Threshold \citep{Campbell2010b} and the Envelope Peaks-Over-Threshold (EPOT) methods \citep{Belenky2011b, Campbell2010}, as well as perturbation-type approaches like the split-time method \cite{Belenky1993, Belenky2010, Belenky2011a}.
	
	Another category of extreme event methodologies are  wave group methods that allow for actual observations of  extreme events. These wave group methods include the Design Loads Generator (DLG) from \cite{Alford2008, Alford2011, Kim2012}, where response amplitude operators (RAO) estimate an extreme value distribution and wave trains are designed to satisfy the estimated distribution. An additional wave group approach is the sequential sampling methodologies from \cite{Mohamad2018, Gong2020} where wave groups are parameterized by their overall length and amplitude. Therefore, knowing the probability of each wave group and predicting the corresponding maximum response enables the development of a probability density function (PDF).   
	
	Another wave group method is the critical wave groups method (CWG), first developed with regular waves in \cite{Themelis2007} and then extended to irregular waves in \cite{Anastopoulos_etal2016, Anastopoulos2016, Anastopoulos2017, Anastopoulos2019}. The underlying idea of the CWG method is that the probability of a response exceeding a threshold is equal to probability of all the pairs of wave groups and ship motion states at the moment of encounter that result in a threshold exceedance. The \emph{critical} wave groups are those that lead to a near-exceedance of the threshold, and therefore any wave group of similar form and ship encounter conditions with larger wave heights also result in an exceedance. Starting from the largest wave in the group with a given height and period, the CWG method utilizes a Markov chain to construct a deterministic wave group based on the most likely successive wave. The majority of the research with the CWG method has considered a single degree-of-freedom (DoF) ordinary differential equation (ODE) model for roll. With an ODE, the encounter conditions can be treated as initial conditions and the simulation of the response due to excitation from the wave groups can be instantiated impulsively. However, if the critical wave group method is to be implemented with higher fidelity hydrodynamic tools or model tests, both the wave groups and encounter conditions must be physically realizable.  A wave group can not impulsively appear and meet the ship with a given initial condition in a wave basin or high-fidelity time-domain numerical simulation. Thus, description of the fluid and body state and its history must be prescribed, which is impractical for computational fluid dynamics (CFD) and impossible in a model testing environment.
	
	\cite{Silva2021oe} addresses the issues with explicitly prescribing encounter conditions and instantaneously starting simulations of the deterministic wave groups by introducing the concept of natural initial conditions. First, the ship response in random seas is simulated with CFD. Then, the deterministic wave groups constructed from Markov chain predictions in the CWG method are embedded into the previously simulated random wave trains, such that the body state of interest occurs at the moment of encountering the deterministic wave group. Embedding the wave group results in a composite wave train that has both an encounter condition and wave group that corresponds to the probability of exceedance calculation developed for CWG. This methodology of forming composite wave trains is applicable to both low and high-fidelity hydrodynamic tools as well as model testing, and allows for repeatable realizations of extreme events. Although the CWG method was successfully implemented with CFD, significant computational cost remains due to the quantity of simulations required to identify critical wave groups for a variety of different encounter conditions and wave group parameters. Therefore, a methodology is required to reduce the total quantity of CFD simulations.
	
	Previous research by \cite{Mohamad2018} and \cite{Gong2020} utilized the concept of sequential sampling and Gaussian Process Regression (GPR) to address the issue of computational expense when predicting statistics of extreme events for marine dynamical systems. By developing a GPR surrogate model of the ship response with fewer simulations and improving the model through an optimization that targets the extremes, they achieved a converged model that exhibits the same statistical behavior of the underlying dynamics with few simulations. However, the GPR models in \cite{Mohamad2018} and \cite{Gong2020} create a mapping between a a wave group parameterization and the maximum response due to that wave group, and do not retain any of the temporal information, thus, understanding the mechanisms leading to extremes is difficult. The present paper aims to retain the temporal response during extreme events by incorporating the methodology developed in \cite{Xu2021, Silva2022aor}, where long short-term memory (LSTM) neural networks are trained to learn the time-accurate ship response due to instantaneous wave elevation. A large component of the implementation of CWG with CFD is the construction of the composite wave trains. The utilization of LSTM allows for the entire ship response to be represented in the surrogate model, rather than simply characterizing the statistics that summarize the particular response ({\it e.g.} maximum). However, the previous studies that utilize a LSTM neural network for ship motion prediction only consider a random nominal wave field \citep{Xu2021, Silva2022aor, Ferrandis2021, DAgostino2022}. The present paper focuses on developing a model that can predict the response time-history due to an excitation from the composite wave trains with embedded deterministic wave groups within random seas. The end result is a trained LSTM neural network that can identify all the critical wave groups rapidly for various response thresholds to calculate the probabilities of exceedance for a particular response. In contrast to previous work, the LSTM neural network model developed in the present paper is trained for extreme motions that exhibit strong nonlinearity.
	
	The objective of the current research is to build off the implementation of the CWG with CFD in \cite{Silva2021oe} and the LSTM framework from \cite{Xu2021} and \cite{Silva2022aor} to develop a new framework that utilizes an LSTM neural network model to reduce the computational burden and provide predictions of the critical wave groups. The improved framework will result in predictions of the probability of exceedance for various response thresholds and a trained neural network model, capable of identifying more critical wave groups to observe with CFD and providing a higher resolution in the probability of exceedance calculations. Additionally, the present paper explores and compares both a general and ensemble modeling approach. The general approach utilizes a single neural network to train all the composite wave trains over the entire parameter range of interest, while the ensemble model approach builds several models where each is responsible for wave groups within a subset of the total parameter space of interest.
	
	The remainder of the current paper is organized as follows. A brief summary of the CWG method and the previously developed framework from \cite{Silva2021oe} is presented, followed by an overview of the considered neural network architecture and methodology. Then, the proposed improved framework with a neural network driven surrogate is detailed along with the two modeling approaches. Finally, the new framework is demonstrated with a case study of a midship section of the Office of Naval Research Tumblehome (ONRT) hull form experiencing extreme roll. The two neural network modeling approaches are compared and the effect of training data quantity on accuracy is explored.
	
	\section*{Critical Wave Groups Method}
	
\begin{figure}[H]
		\centering
		\includegraphics[width=0.6\textwidth]{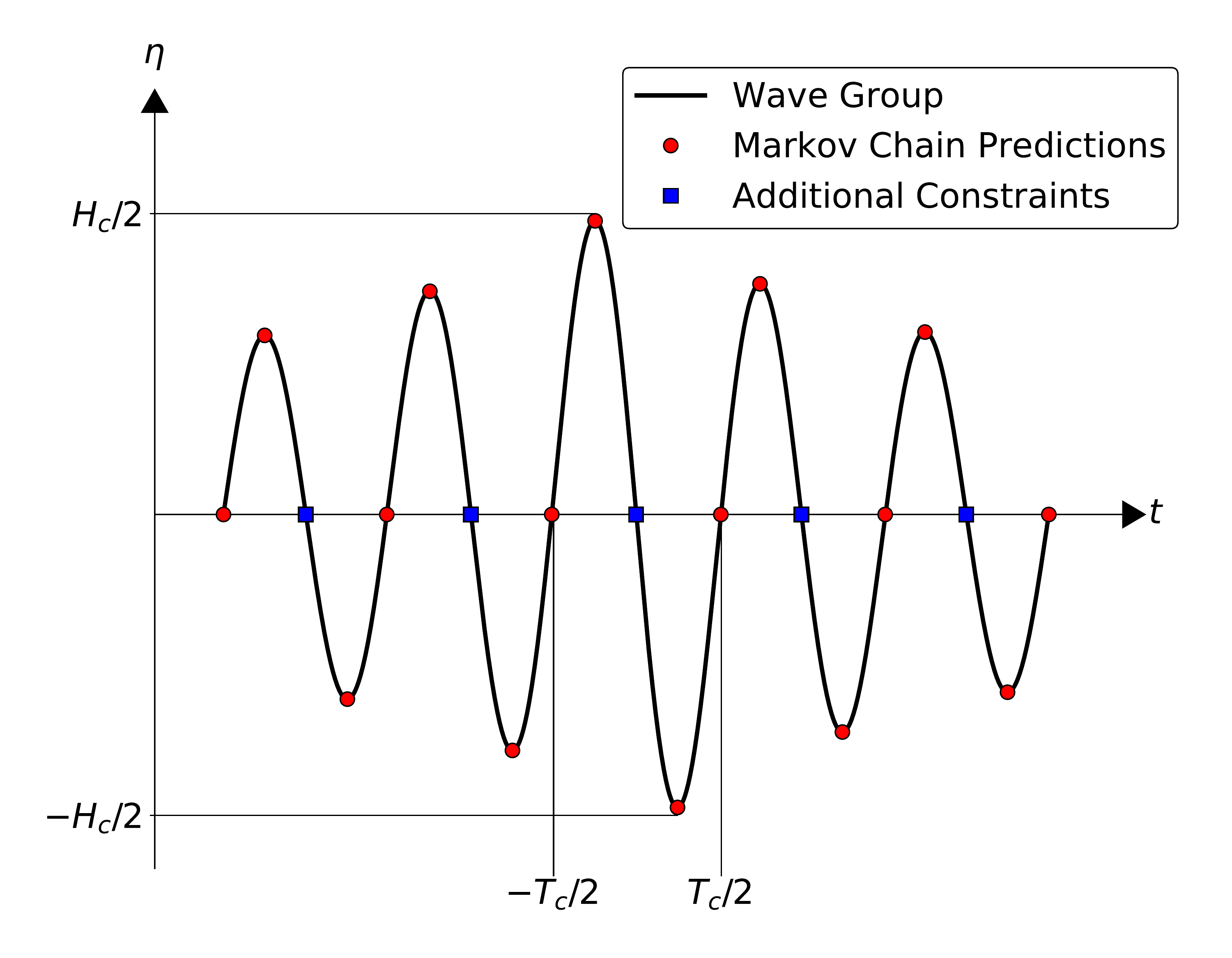}
		\caption{Markov chain construction of wave groups and additional geometric constraints.}
		\label{fig:markovChain}
	\end{figure}

	The present paper derives directly from the CWG method developed in \cite{Themelis2007, Anastopoulos_etal2016, Anastopoulos2016, Anastopoulos2017, Anastopoulos2019}. The main idea of the CWG method is to identify wave groups for a selected set of ship states at the moment of encountering the wave group  (encounter conditions) that lead to a near-exceedance of a specified response threshold. In previous research with ODE models of roll, these encounter conditions are typically referred to as initial conditions. Wave groups in the CWG method are constructed systematically with Markov chains and the statistical relationship between successive wave heights and periods. The memoryless property of Markov chains allows for predictions of the most likely successive waves, given the height and period of the current wave.  Therefore, wave groups containing $j$ waves can be constructed solely by prescribing the height and period of the largest wave of the group ($H_c$, $T_c$). Fig.~\ref{fig:markovChain} shows a given wave group with the heights and periods predicted through the Markov Chain predictions. Additional constraints such as the location of zero crossing and size of crest relative to wave height are assumed in accordance with \cite{Anastopoulos2019} in order to produce a continuous representation of a wave group with a Fourier basis through trigonometric interpolation \citep{Nathan1975}. 
	
	Without considering encounter conditions, categorizing the deterministic wave groups by $H_c$, $T_c$ and $j$, critical wave groups for a given  $T_c$ and $j$ can be found by only varying the value of $H_c$. An infinite number of values of $T_c$  are possible. Therefore, all wave periods are discretized into $m$ intervals of length $\Delta T$, where $T_c$ is representative of the wave periods in that interval. A $\Delta T$ of 1 is utilized for the present paper in accordance with the findings of \cite{Anastopoulos2019}. When considering the effect of encounter conditions on identifying critical wave groups, the possible values must also be discretized into $k$ components that are representative of the intervals with which they correspond. A single critical wave group can be identified for wave groups where the period of the largest wave is $T_c$, the run length is $j$, and the encounter condition is $ec_k$, by varying the height of the largest wave in the group. These critical wave groups can be combined into the expression in Eqn.~\eqref{eq:probexceed_param}, where the probability of the response $\phi$ exceeding $\phi_{\textrm{crit}}$ is expressed as the combination of the probability of observing groups larger than the critical wave groups $wg_{m,j}^{(k)}$ and the probability of the encounter conditions $ec_k$ \citep{Anastopoulos2019}.
	
	\begin{equation}
	p \left[ \phi > \phi_{\textrm{crit}} \right] = 
	\sum\limits_{k}{} 
	\sum\limits_{m}{}
	\left(	
	1-
	\prod_{j}\left( 1-p \left[ wg_{m,j}^{(k)} \right]\right)
	\right)
	\times
	p \left[ ec_k \right]
	\label{eq:probexceed_param}
	\end{equation}
	
	The probability of encountering groups larger than the critical wave groups $wg_{m,j}^{(k)}$ is calculated by first identifying the values of $H_c$ that correspond to the near-exceedance of a specified threshold for all the different combinations of $T_c$, $j$, and $ec_k$. Then, the probability of exceeding each of those critical wave groups that are uniquely described by $H_c$, $T_c$, and $j$ is calculated with Eqn.~\eqref{eq:probwg}, where the probability of encountering a wave group larger than $wg_{m,j}^{(k)}$ for the given period range $m$, run length $j$, and initial condition $k$ is equal to the joint probability of encountering a wave group with heights larger than each wave height in the critical group, $\mathbf{h}_{cr}^{(k)}$, and the periods of each wave in the group being within the wave period range $T_{cr,m}$. The probability of the encounter conditions $p\left[ec_k\right]$ can be found by sampling a probability distribution of the quantities of interest through simulations in random irregular waves. Detailed descriptions of Eqn.~\eqref{eq:probexceed_param} and \eqref{eq:probwg} can be found in \cite{Anastopoulos2019}.
	
	\begin{equation}
	p \left[ wg_{m,j}^{(k)} \right] \ = \ 
	p \left[ \mathbf{H}_j > \mathbf{h}_{cr}^{(k)},\mathbf{T}_j \in T_{cr,m} \right]
	\label{eq:probwg}
	\end{equation}
	
	\section*{CWG-CFD Framework}
	
\begin{figure}[H]
		\centering
		\includegraphics[width=0.5\textwidth]{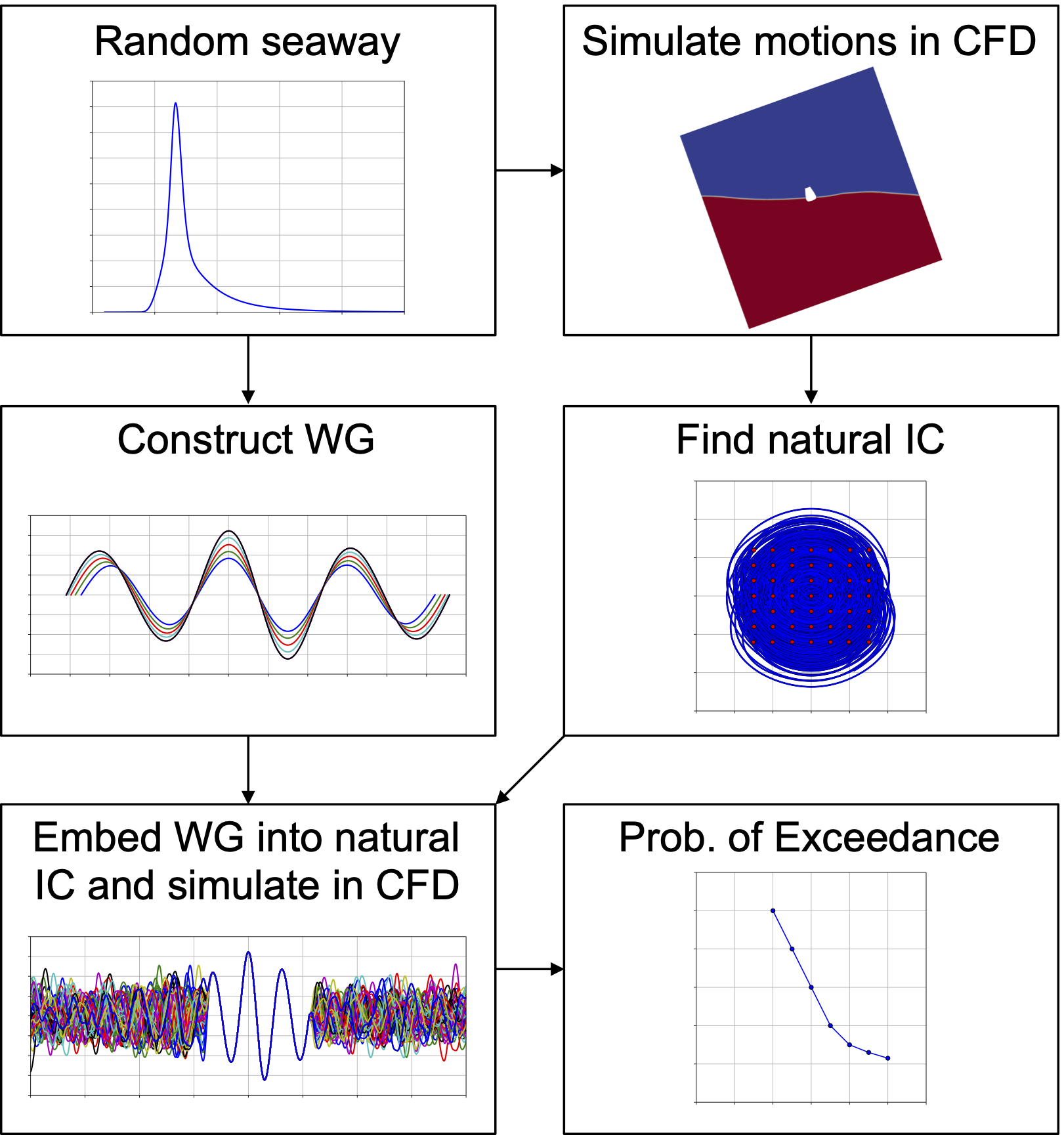}
		\caption{Flow chart of CWG-CFD framework from \cite{Silva2021oe}.}
		\label{fig:flowChartCWGCFD}
	\end{figure}

	The implementation of the CWG method with CFD (CWG-CFD) was first introduced in \cite{Silva2021oe} and an overview of the method is shown in Fig.~\ref{fig:flowChartCWGCFD}. The framework begins with a selection of a seaway of interest and the spectrum is sampled to produce random irregular wave time histories to both simulate in CFD and develop the statistical relationships between successive wave heights and periods. The successive wave statistics are then utilized with Markov chains to construct the deterministic wave groups necessary for the CWG method. The CFD simulations in random waves are used to develop the probability distribution of encounter conditions, as well as identify sequences of waves that lead to motion states of interest for prescribing different ship motion states at the moment of wave group encounter, referred to as natural initial conditions. The deterministic wave groups and natural initial conditions are combined into composite wave trains that are then simulated with CFD. The resulting maximum responses from the simulated composite wave trains are then utilized to identify the critical wave groups for various response thresholds, which are then leveraged in the probability of exceedance calculation. To produce physically realizable wave groups with specified ship motion states at the moment of encounter, \cite{Silva2021oe} introduced the natural initial condition concept. Simulations of a ship in random irregular waves are used to identify motion states of interest and the series of waves they originated from. Then,  deterministic wave groups constructed from the CWG method are embedded into the wave train, such that the motion states of interest occur at the start of the wave group. Embedding the wave group in this manner results in a single and repeatable composite wave train that can be simulated, preserving the integrity of the CWG methodology with a deterministic wave group and prescribed encounter state. The natural initial condition concept can be also extended to other hydrodynamic simulation tools and model tests, as well as different wave group probabilistic frameworks where wave groups are constructed and varying the encounter state when encountering the wave group is desired.
	
\begin{figure}[H]
		\centering
		\includegraphics[width=0.6\textwidth]{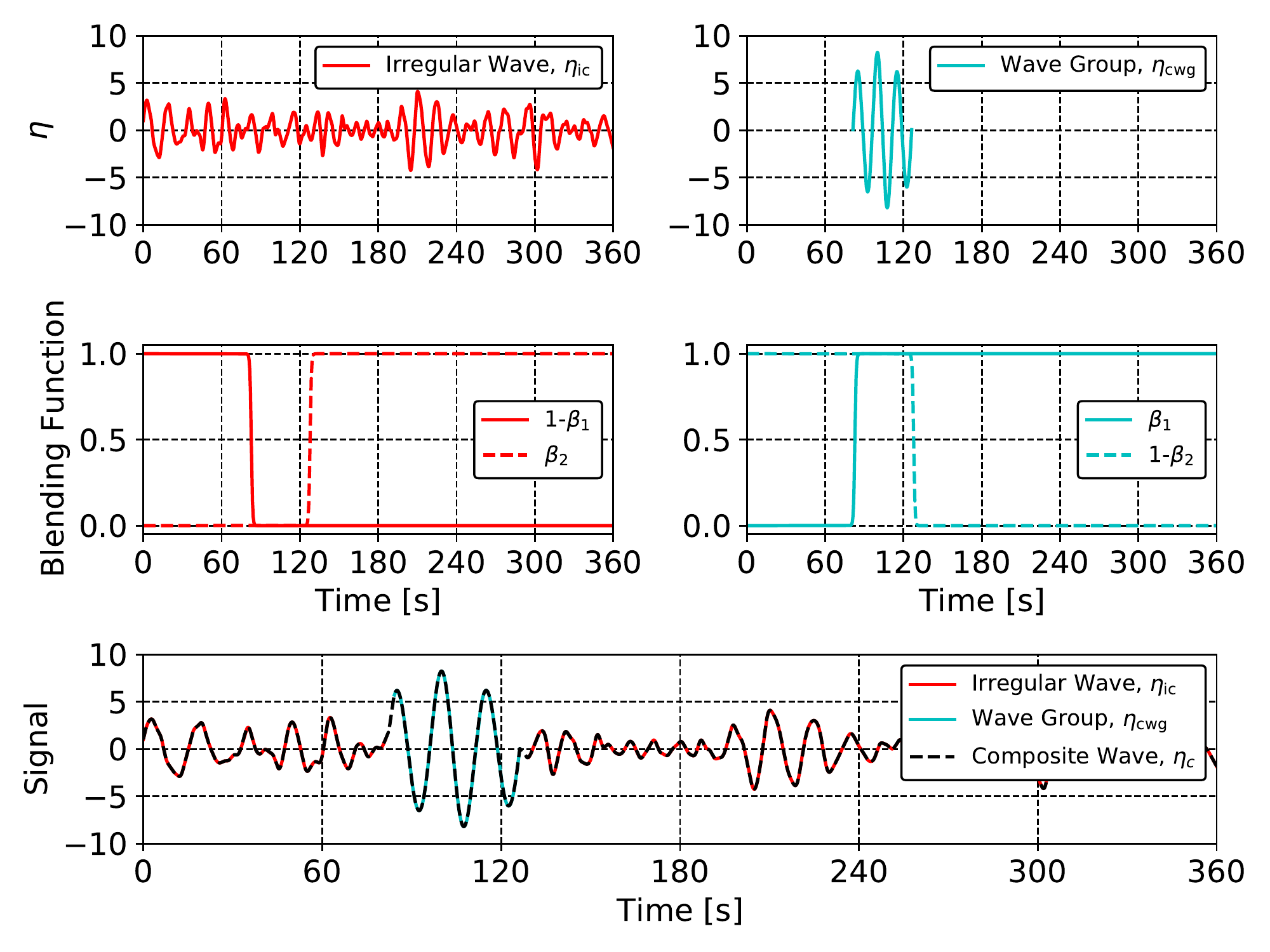}
		\caption{Formation of a composite wave by embedding a deterministic wave group into an irregular wave train.}
		\label{fig:blend}
	\end{figure}

		Fig.~\ref{fig:blend} shows the blending process for embedding the wave group, $\eta_{\rm{cwg}}\left(\mathbf{x},t\right)$, into an irregular wave train, $\eta_{\rm{ic}}\left(\mathbf{x},t\right)$, to create a single composite wave train, $\eta_{\rm c}\left(\mathbf{x},t\right)$. The composite wave train is described by:
	\begin{equation}
	\eta_{\rm c} = \left(1 - \beta_2\right)\left[(1-\beta_1)\eta_{\rm ic} + \beta_1\eta_{\rm cwg}\right]
	+ \beta_2\eta_{\rm ic}
	\end{equation}
	
	\noindent where each blending function is defined as:
	\begin{equation}
	\beta \ = \frac{1}{2}\left(1 + \tanh\left( \frac{t \ - \ t_b}{t_o} \right) \right)
	\label{eq:blend}
	\end{equation}
	
	The functions $\beta_1$ and $\beta_2$ correspond to the blending at the beginning and end of the wave group, respectively. Fig.~\ref{fig:blend} shows the blending process for embedding the wave group into an irregular wave train to create a single composite wave train. The two parameters in Eqn.~\eqref{eq:blend},  $t_b$ and $t_o$, control time shift and scale of the overlap between $\eta_{\rm{cwg}}\left(\mathbf{x},t\right)$ and $\eta_{\rm{ic}}\left(\mathbf{x},t\right)$. The time shift $t_b$ is selected to be $T_p$/10 s from the start or end of the wave group to enforce that 95\% of the composite wave train is the irregular wave train at $t$ = $t_b$. The time scale $t_o$ is selected with Eqn.~\eqref{eq:to}, where the factor of 0.9 corresponds to approximately 95\% of the first signal at the beginning of the blending interval and 95\% of the second signal at the end of the blending interval, and $T_p$ is the peak modal period of the seaway. A composite wave shares the same $t_o$ for $\beta_1$ and $\beta_2$, but the value of $t_b$ depends on the beginning and end of the wave group.
	
	\begin{flalign}
	t_o \ = \ \frac{T_p}{10 \cdot \rm tanh^{-1}(0.9)}
	\label{eq:to}
	\end{flalign}
	
\begin{figure}[H]
		\centering
		\includegraphics[width=0.5\textwidth]{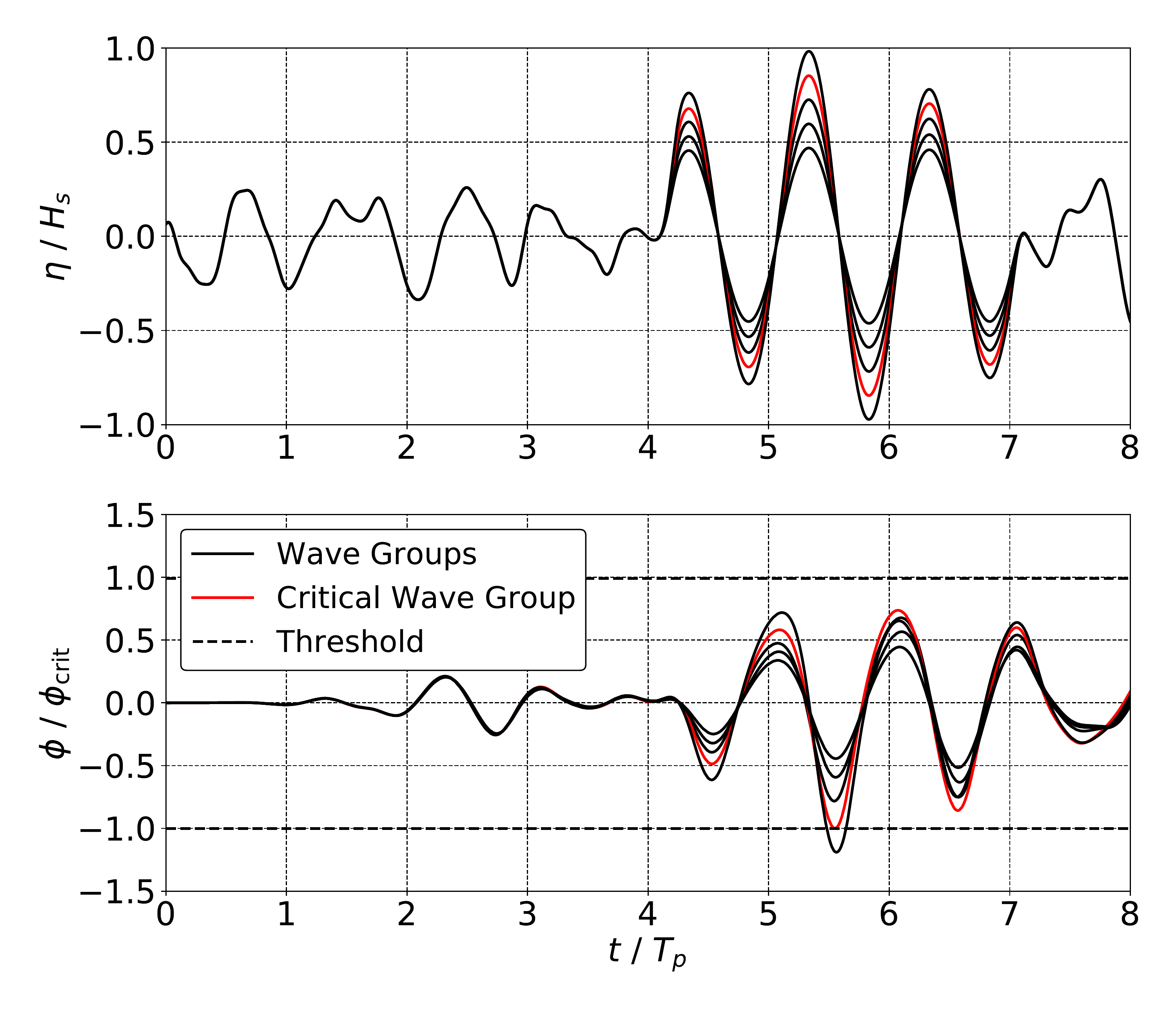}
		\caption{Identification of a critical wave group for a given set of wave groups with similar shapes.}
		\label{fig:crit}
	\end{figure}

	Fig.~\ref{fig:crit} demonstrates how a critical wave group is identified for a given response $\phi$ exceeding a threshold $\phi_{\textrm{crit}}$. Each curve in Fig.~\ref{fig:crit} represents the wave elevation, $\eta$, and corresponding response, $\phi$, time histories for a given set of composite waves with identical encounter conditions and wave groups with the same period of the largest wave $T_c$  and run length $j$. Varying only the height of the largest wave in the group, a critical group can be found that results in a near exceedance of $\phi_{\textrm{crit}}$. At each desired threshold, encounter condition, wave period range, and run length, the procedure shown in Fig.~\ref{fig:crit} can be performed for a series of values for the height of the largest wave in the group to identify all of the critical wave groups. The corresponding probabilities of each of those critical wave groups and encounter conditions are combined with Eqn.~\ref{fig:flowChartCWGCFD} to calculate the probability of exceedance at a given  threshold.

	\section*{Neural Network Model}
	
	The present paper relies heavily on the methodology developed in \cite{Xu2021} and \cite{Silva2022aor} for constructing a system identification model that represents the dynamical response of a vessel in waves with LSTM neural networks. Ship motions in waves are an example of a causal dynamic system. Thus, their output not only depends on the current force excitation (eg. waves), but also the previous excitation as well. The output of a discrete dynamical system $y_t$ can be described by: 
	\begin{equation}
	y_t = f(x_t, x_{t-1}, x_{t-2}, \cdots)
	\label{eq:reg_output}
	\end{equation}
	where $f$ is a mapping function and $x_t$ corresponds to the input at time index $t$. Eqn.~\eqref{eq:reg_output} demonstrates that the output state $y_t$ not only depends on the current input $x_t$, but also previous values ($x_{t-1}, x_{t-2},...$). The overall goal of training the model is to develop the best nonlinear mapping $f$ that describes the underlying dynamics. In the current methodology, the input $x_t$ corresponds to a description wave time-history at multiple wave probes in the CFD domain, the output $y_t$ is the response of the vessel, and the mapping is found through the construction and training of an LSTM neural network model.    
	
\begin{figure}[H]
		\centering
		\includegraphics[width=0.5\textwidth]{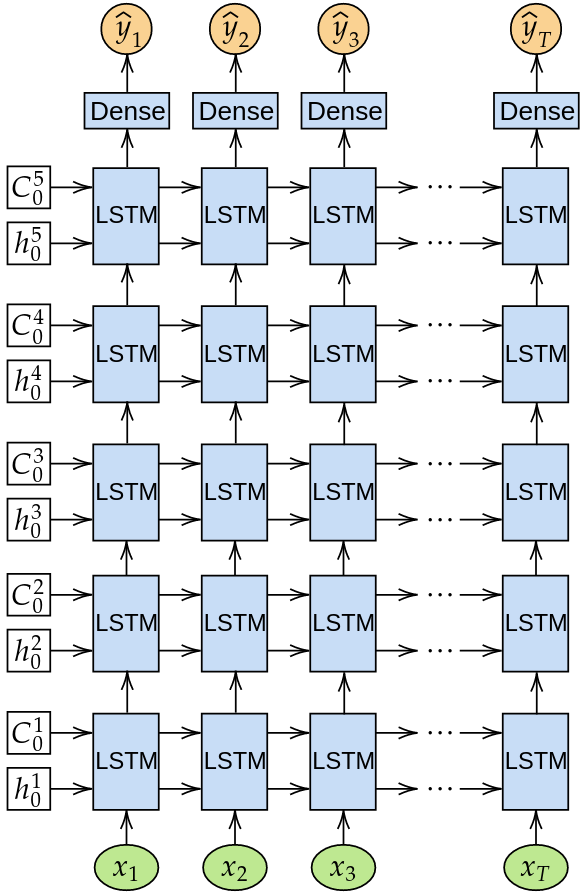}
		\caption{Neural network architecture from \cite{Xu2021}}
		\label{fig:dre-arch}
	\end{figure}

	Fig.~\ref{fig:dre-arch} displays an example neural network architecture with five LSTM layers, followed by a dense layer. Inputs and outputs are denoted as $x_t$ and $y_t$ respectively, where $t$ is the time step index that ranges from 1 to $T$. $C_t^n$ and $h_t^n$ correspond to state and output of LSTM cell $n$ at time index $t$. The stacking of LSTM layers allows for the output of the previous layers to be used as the input for the next layer. Generally, adding layers to the neural network model architecture allows for a greater level of abstraction within the trained model, which allows for generalized predictions of input scenarios that are not considered during the model training. The dense layer in Fig.~\ref{fig:dre-arch} employs a linear activation function and receives the last of the LSTM layers as input, and outputs the final result of the neural network model. The model architecture in this paper is implemented with the toolbox Keras \citep{Chollet2015} with Tensorflow \citep{Abadi2015} as its backend.
	
	The first step of building the model is to identify the input and corresponding outputs and split the data into training and testing subsets. The model is trained with the training dataset, while the test set is independent of the training and is utilized to validate the model. The inputs and outputs are then standardized such that the mean of each is zero and the standard deviation is one. The scaling of the features avoids instances of one quantity dominating the training process. Next, the model enters an optimization process to minimize the difference between the model predictions and the training data, which is then quantified with a loss function, or metric of how close the model predictions are to the data. The loss function utilized in the present paper is the mean-squared error averaged over all time steps shown in Eqn.~\eqref{eq:loss}, where $\hat{y}$ is the prediction of the output sequence and $y$ is the true value.
	
	\begin{equation}
	L(\hat{y}, y) = \frac{1}{T}\sum_{t=1}^T (\hat{y}_t - y_t)^2
	\label{eq:loss}
	\end{equation}
	
	The training of the model at every iteration starts with forward propagation, where the model computes from the input layer towards the output layer with the current model parameters. Then, the loss is computed with outputs from the current model. After, backward propagation computes the loss derivative with respect to each of the parameters within the model from the output layer to the input layer. Based on the loss derivatives, the model parameters are updated to minimize the loss function and the process is repeated. The case studies presented in this paper train the neural networks with an Adam optimizer \citep{kingma2014adam}.
	
	In the current paper, the input to the neural network is the wave elevation time history at 27 wave probes in the CFD domain in accordance with \cite{Silva2022aor} and the output is the ship heave and roll temporal response. The developed models all utilize two LSTM layers followed by a dense layer, with each LSTM layer containing 50 cells. This model architecture follows the ship motion example considered in \cite{Xu2021}. Overall, the LSTM approach can produce a time-accurate representation of the ship response, where the time histories for heave and roll motion are compared for a composite wave train that is not included in the training dataset for the neural-network model. Generating the temporal response of the motions for a given composite wave train provides insight into how extremes occur as opposed to approaches taken in \cite{Mohamad2018} and \cite{Gong2020}, which only consider the maximum value of the response in their surrogate models. The ability to produce the response time histories with the LSTM neural network becomes much more useful when considering complex 6-DoF motions.
	
	The current paper also implements the Monte Carlo Dropout approach developed by \cite{Gal2016a, Gal2016b} to quantify the uncertainty in the LSTM neural network model predictions. Dropout is typically employed as a regularization technique during training to avoid overfitting a model by randomly excluding a portion of the neural network. The Monte Carlo Dropout approach applies the dropout to the prediction as well and provides an ensemble of stochastic predictions, which can be converted into a mean prediction with an uncertainty estimate. The present methodology implements the Monte Carlo Dropout method by adding a dropout layer after each LSTM layer in Fig.~\ref{fig:dre-arch}.

	\section*{CWG-CFD-LSTM Framework}
	
\begin{figure}[H]
		\centering
		\includegraphics[width=0.5\textwidth]{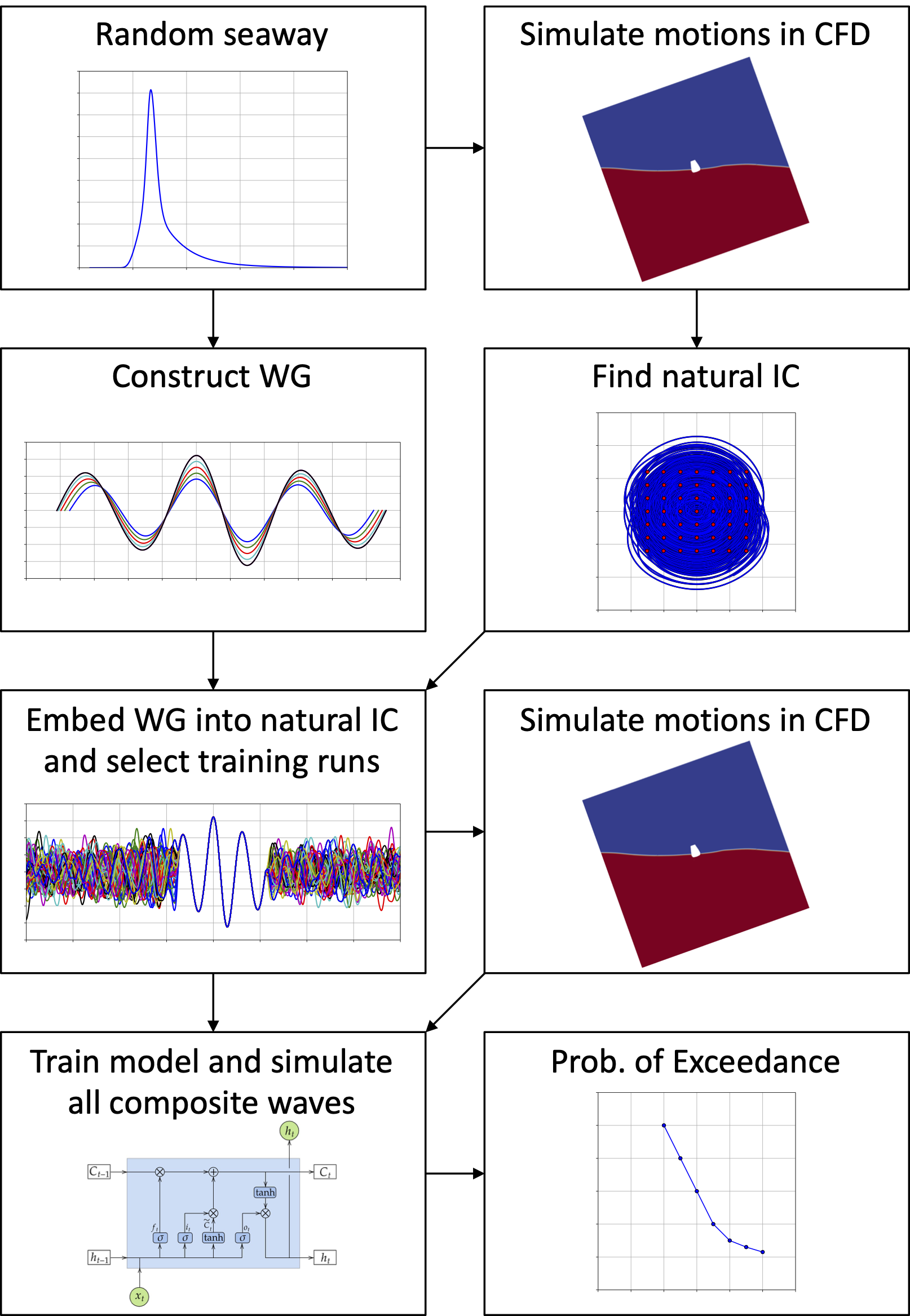}
		\caption{Flow chart of proposed CWG-CFD-LSTM framework.}
		\label{fig:flow-cwg-cfd-lstm}
	\end{figure}

	The CWG-CFD framework developed in \cite{Silva2021oe} is capable of producing predictions of extreme events and the present paper aims to recover the same quantitative calculations, but at a fraction of the computational cost with the addition of LSTM neural networks. The CWG-CFD-LSTM approach outlined in Fig.~\ref{fig:flow-cwg-cfd-lstm} is identical to the CWG-CFD framework from \cite{Silva2021oe}, up to the point of constructing the composite wave trains with the natural initial conditions and embedded deterministic wave groups. The CWG-CFD-LSTM differs starting with the random selection of a specified quantity of composite wave trains. Then, the ship response due to each of those composite wave trains is simulated with CFD. An LSTM neural network model (or multiple) is then trained with the CFD simulations of ship response due to the composite wave trains. After training, the different neural network models are then employed to simulate all the remaining composite wave trains and identify the critical wave groups. The critical wave groups, encounter conditions, and their respective probabilities are then considered with Eqn.~\eqref{eq:probexceed_param} to calculate the probability of exceedance.
	
	The current paper considers both a general and an ensemble modeling approach. The general approach considers a single neural network model that is trained with randomly selected composite wave trains. The ensemble approach utilizes several models, each responsible for composite wave trains, with a specified period of the largest wave $T_c$ and run length $j$. Both modeling approaches are explored because although the general approach allows for more training runs per model. When compared to an ensemble approach in terms of total training runs required, it must predict the dynamics over a larger parameter range.  The ensemble model approach only considers wave groups within a smaller subset of the total parameter range. Therefore, the ensemble model approach only needs differentiate between the height of the largest wave in the group and the various encounter conditions, which can provide added accuracy and faster convergence with respect to training data quantity. 
		
	\section*{Case Study}
	
\begin{figure}[H]
		\centering
		\includegraphics[width=0.23\textwidth]{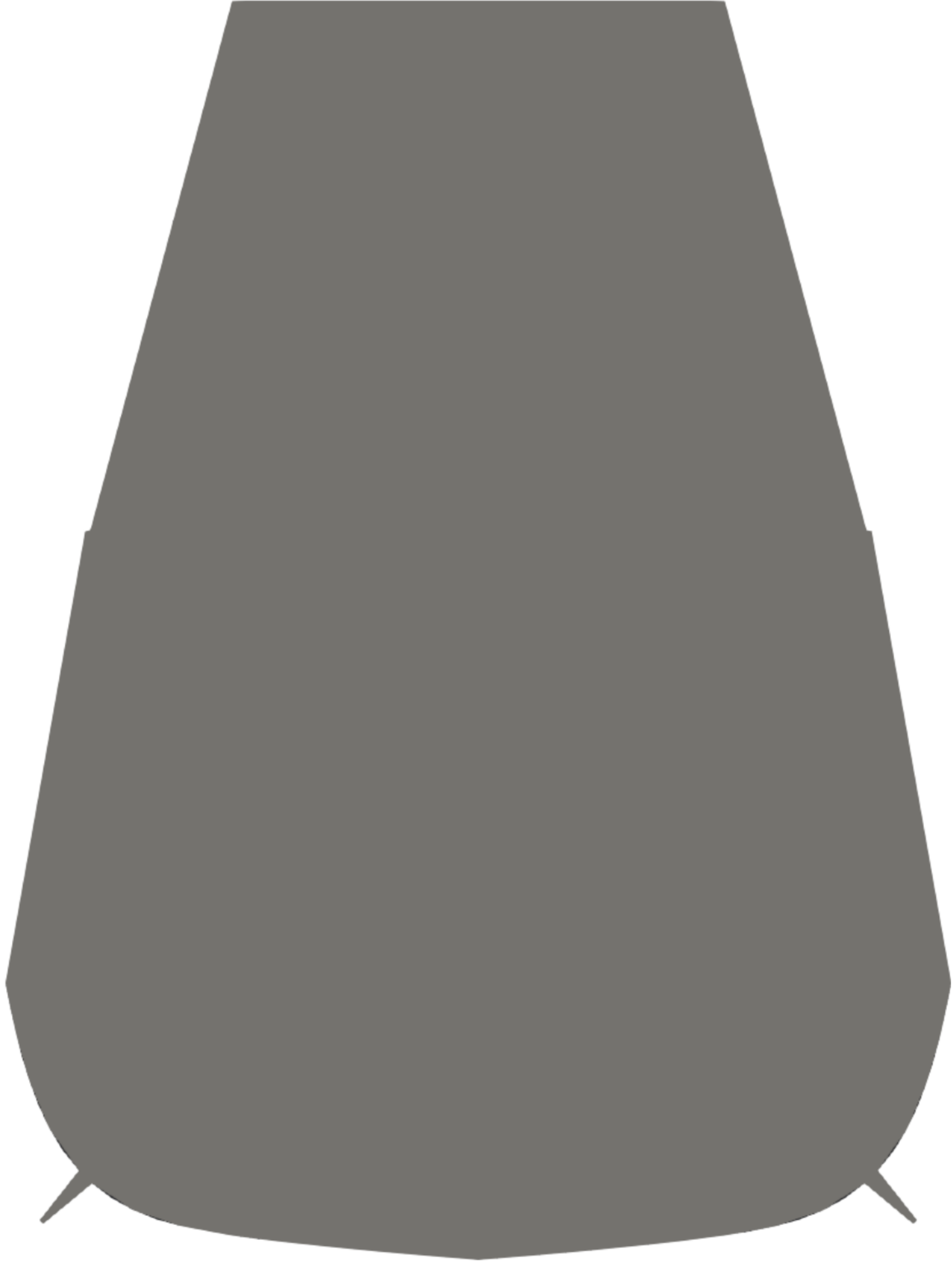}\hspace{1cm}\includegraphics[width=0.31725\textwidth]{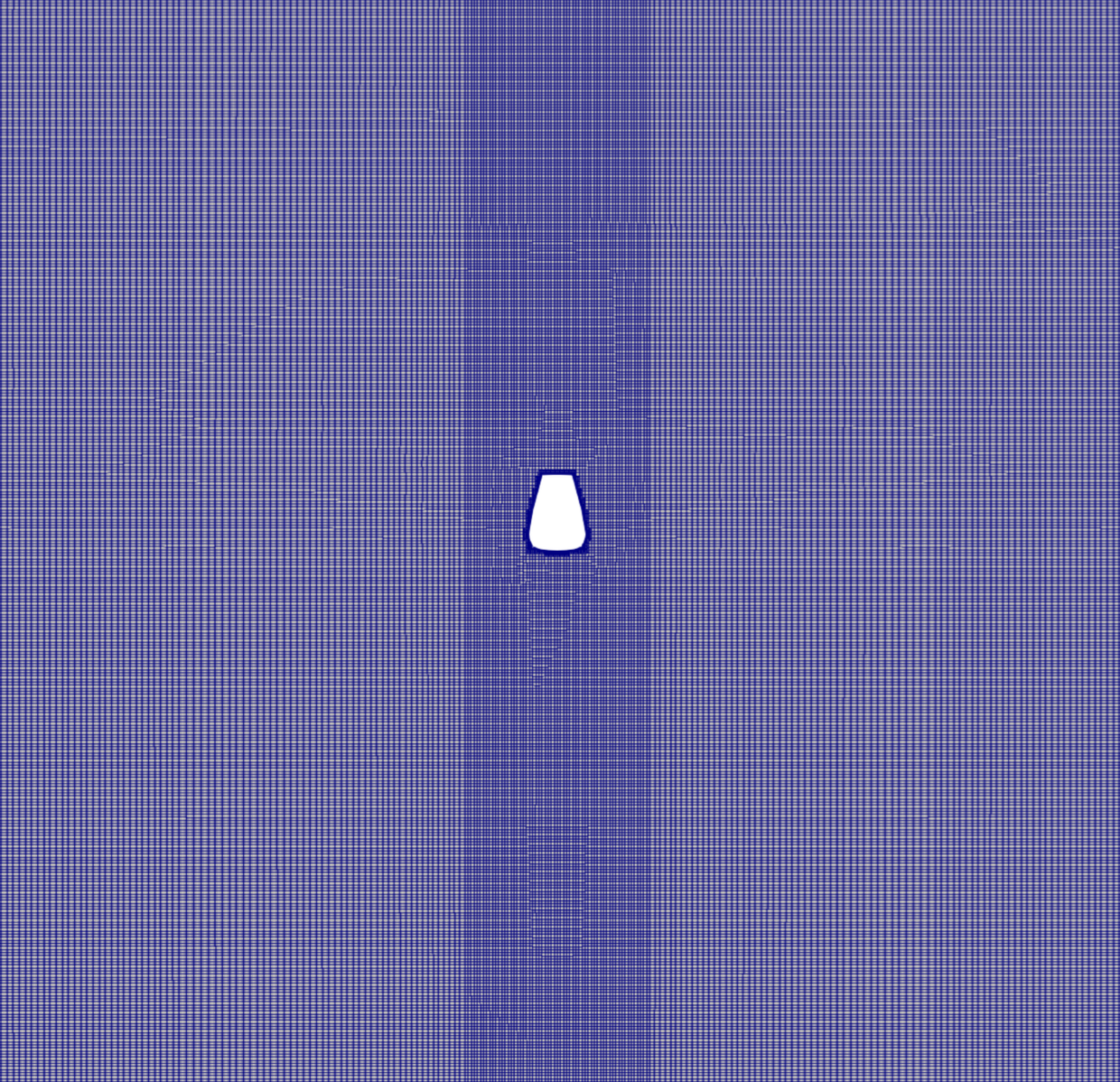}
		\caption{2-D ONRT midship section geometry and computational mesh.}
		\label{fig:hull}
	\end{figure}

	The proposed CWG-CFD-LSTM framework is demonstrated with the same case study considered in the CWG-CFD framework developed in \cite{Silva2021oe} of a two-dimensional (2-D) midship section of the ONRT geometry \citep{Bishop2005} shown in Fig.~\ref{fig:hull}, with hull and fluid properties shown in Table \ref{tab:load}. The case study uses the open-source toolkit OpenFOAM\textregistered{}, to simulate the ship motion response due to nonlinear generated seaways with customized CFD solvers and libraries developed by the Computational Ship Hydrodynamics Laboratory (CSHL) at The University of Michigan \citep{Filip2017, Piro2013}. The response of a 2-D ONRT midship section in waves demonstrates the ability of the LSTM models to represent the underlying nonlinear response simulated with CFD and showcases how the presented approach can produce similar extreme statistics with a reduction in the computational cost, in comparison to a purely CFD-driven CWG framework. The case study uses a JONSWAP spectrum \citep{Hasselmann1973} with a peak enhancement factor $\gamma= 3.3$, a significant wave height $H_s = 7.5$~m, and a peak modal period $T_p = 15$~s,  corresponding to a Sea State 7 \citep{NATO1983}. The 2-D midship section is only permitted to heave and roll, and is constrained in the other DoF with waves traveling from beam seas. The considered case study is setup to predict the extreme roll of the midship section and uses roll and roll velocity as encounter conditions.

	\begin{table}[H]
		\caption{Loading condition and fluid properties of 2-D ONRT midship section.}
		\begin{center}
			\label{tab:load}
			\begin{tabular}{l | c | r}
				\hline
				Properties & Units & Value \\
				\hline
				Draft, $T$		&	m	&	5.5	\\
				Beam, $B$			&	m	&	18.8	\\
				Roll Gyradius	&	m	&	7.118	\\
				Vertical Center of Gravity, $KG$ (ABL)	&	m	&	7.881	\\
				Transverse Metacentric Height, $GMT$		&	m	&	1.5	\\	
				Density of Water, $\rho_w$	&	kg/m$^\text{3}$	&	1000	\\	
				Density of Air, $\rho_a$	&	kg/m$^\text{3}$	&	1	\\	
				Kinematic Viscosity of Water, $\nu_w$	&	m$^\text{2}$/s	&	1e-06	\\	
				Kinematic Viscosity of Air, $\nu_a$		&	m$^\text{2}$/s	&	1.48e-05	\\	
				\hline
			\end{tabular}
		\end{center}
	\end{table}
	
	The training matrix, neural network architecture, and hyper-parameters for the case study are presented in Table~\ref{tab:trainmatrix}. The CWG-CFD-LSTM framework is evaluated with training datasets of 50, 100, 200, or 400 training runs for both the ensemble and general neural network modeling approaches. The training dataset is identical between the general and ensemble modeling approaches, and the smaller sized training datasets are a subset of the larger training runs. For example, the models with 100 total training runs utilize the same 50 runs as the models trained with only 50 training runs. For each total quantity level of training data, the runs are segregated equally across the 10 $T_c$ and $j$ pairs from \cite{Silva2021oe}. For each $T_c$ and $j$ pair, training runs are selected randomly in terms of $H_c$ and the encounter conditions.  For the ensemble approach, a separate model is constructed for each of the $T_c$ and $j$ pairs. The general approach utilizes all of the training data as the ensemble model approach, but only builds a single model. For example, the ensemble model approach for 400 total training runs contains 40 runs for each of the 10 $T_c$ and $j$, while the general approach would train off the same 400 runs. The same is true for other training dataset sizes. This breakdown of training data ensures that there is not any bias in the training, and both modeling approaches have the same information available. Each model utilizes the same architecture and training methodology and is evaluated against 25,100 validation runs that correspond to all of CFD simulations required to calculate the probability of exceedance in \cite{Silva2021oe} for both the $L_2$ (formulated as the root mean squared error) and $L_{\infty}$ error described in Eq.~\eqref{eq:L2error} and \eqref{eq:Linferror} respectively.

	\begin{equation}
	L_2(y, \hat{y}) = \sqrt{L(y,\hat{y})} = \sqrt{\frac{1}{T}\sum_{i=1}^T (y_i - \hat{y}_i)^2}
	\label{eq:L2error}
	\end{equation}
	
	\begin{equation}
	L_\infty(y, \hat{y}) = \max_{i=1, \cdots, T} |y_i - \hat{y}_i|
	\label{eq:Linferror}
	\end{equation}
	
\begin{figure}[H]
		\centering
		\includegraphics[width=.5\linewidth]{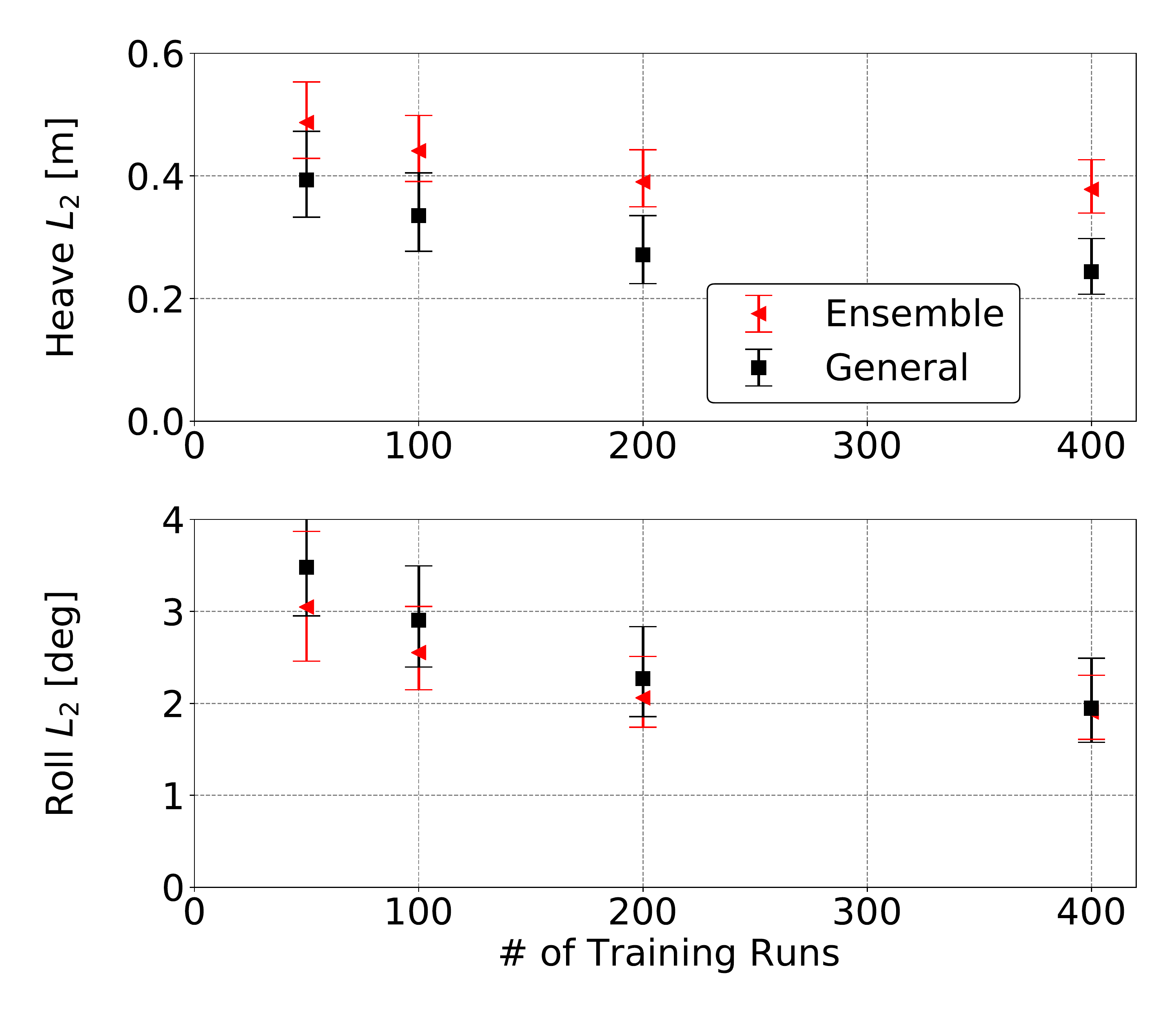}  
		\caption{Comparison of $L_2$ error for heave and roll.}
		\label{fig:L2}
	\end{figure}

\begin{figure}[H]
		\centering
		\includegraphics[width=.5\linewidth]{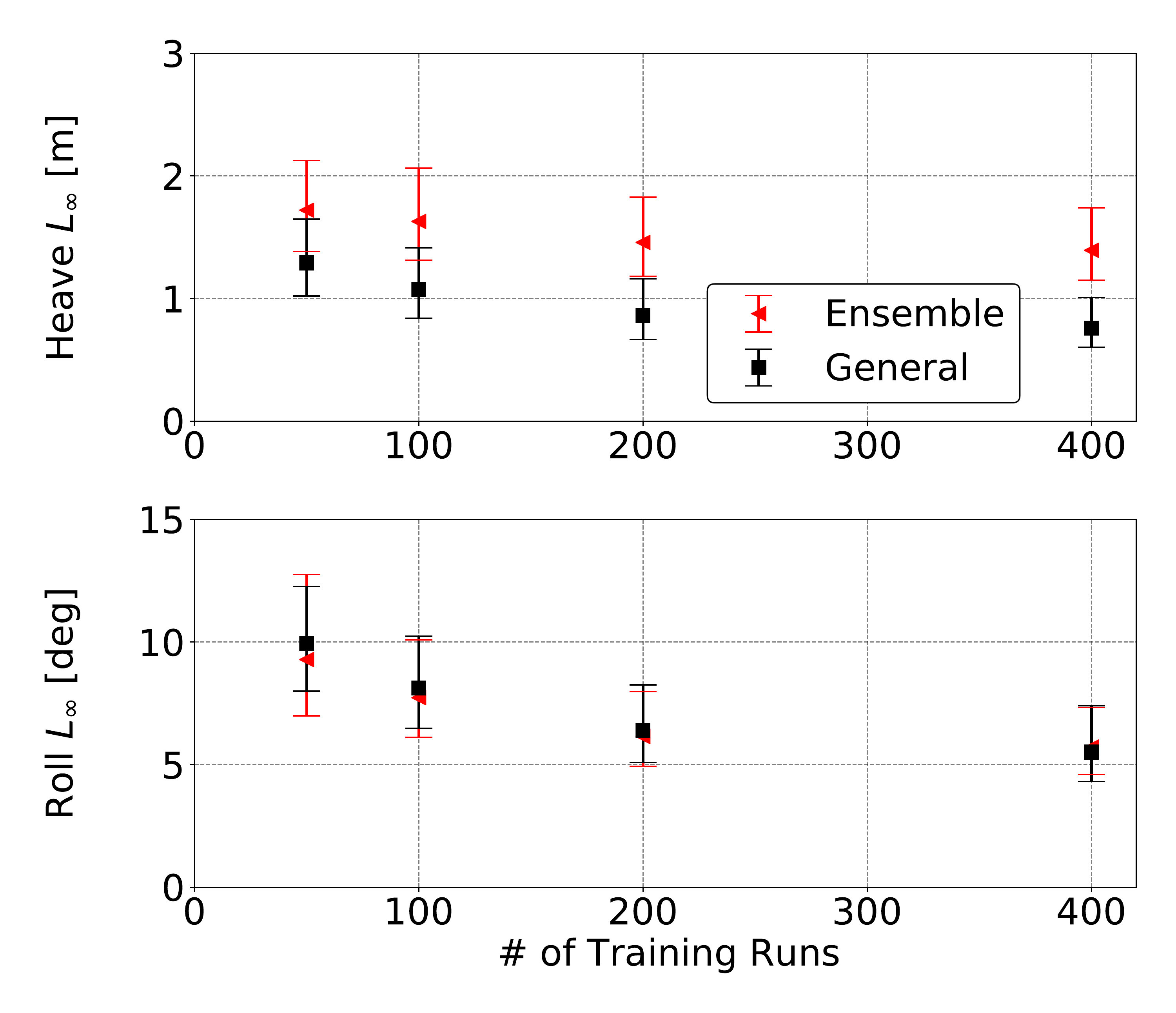}  
		\caption{Comparison of $L_\infty$ error for heave and roll.}
		\label{fig:Linf}
	\end{figure}

	Fig.~\ref{fig:L2} and \ref{fig:Linf} compare the $L_2$ and $L_\infty$ error respectively for heave and roll utilizing both the ensemble and general approaches with various quantities of training data. The error calculation in Eq.~\eqref{eq:L2error} and \eqref{eq:Linferror} are performed for each of the validation runs. The triangle and rectangle markers in Fig.~\ref{fig:L2} and \ref{fig:Linf} correspond to the median error and the error bars denote the 25$^{\textrm{th}}$ and 75$^{\textrm{th}}$ percentiles. For both the general and ensemble approach, the $L_2$ and $L_\infty$ error for heave and roll decreases as the quantity of training data increases. Additionally, the size of the error bars for the $L_2$ and $L_\infty$ error decreases as the quantity of training data increases. Overall, the general approach provides better predictions for heave, but the roll predictions are similar between the two modeling approaches.
		
	\begin{table}[H]
	\caption{Training matrix, neural network architecture, and hyper-parameters for the case study.}
	\begin{center}
		\label{tab:trainmatrix}
		\begin{tabular}{l | l}
			\hline
			Properties  & Value \\
			\hline
			Total Training Runs			&	50, 100, 200, 400	\\
			Training Runs per Model (Ensemble)			&	5, 10, 20, 40	\\
			Total Validation Runs			&	25,100	\\
			Time Steps per Run			&	200	\\
			Units per Layer			&	50	\\
			Layers			&	2	\\
			Dropout			&	0.1	\\
			Learning Rate			&	0.001	\\
			Epochs			&	2,000	\\
			\hline	
		\end{tabular}
	\end{center}
    \end{table}

\begin{figure}[H]
		\centering
		\includegraphics[width=.5\linewidth]{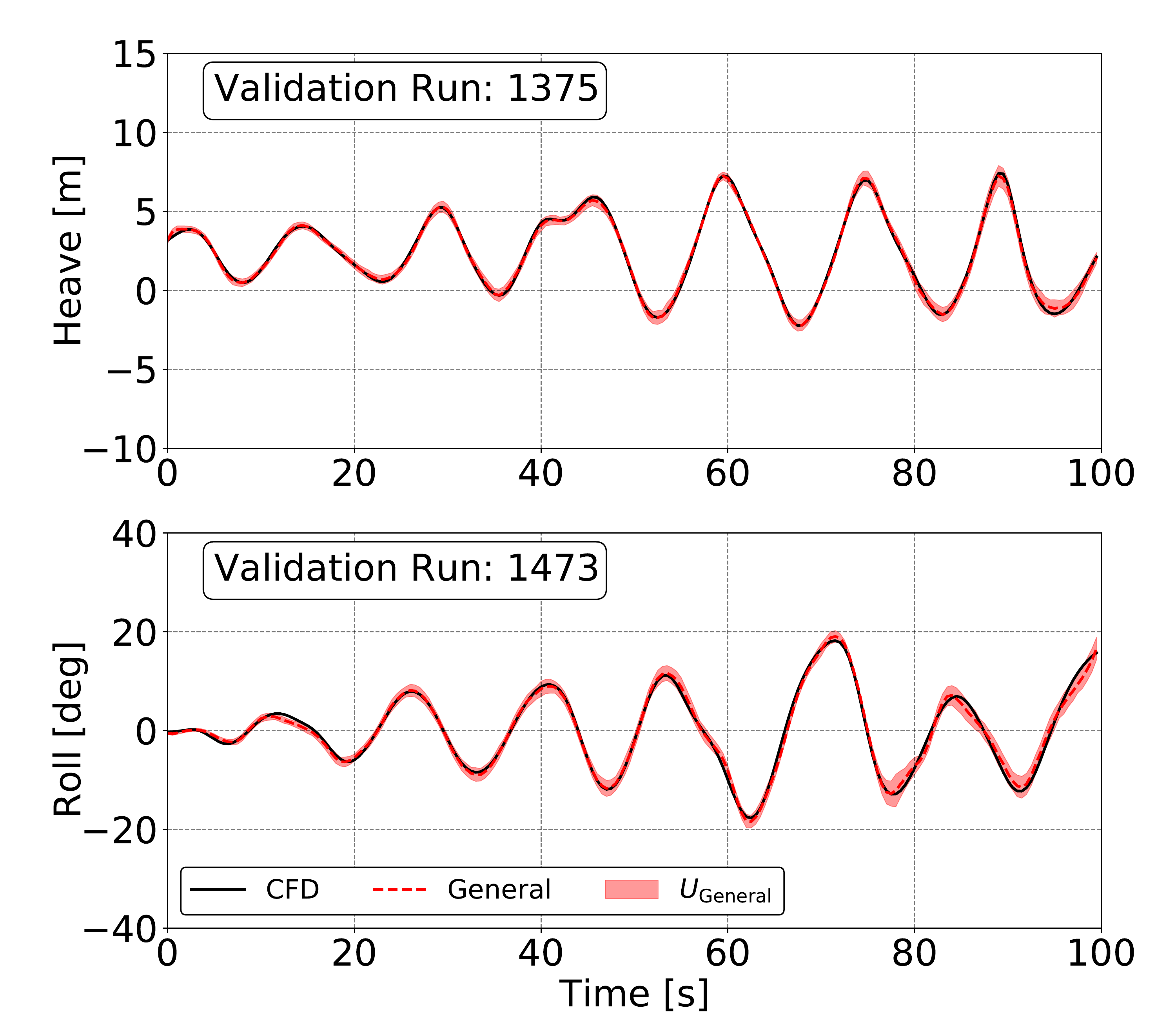}  
		\caption{Comparison of time-histories with the smallest $L_\infty$ error for heave and roll with a model trained with 400 simulations.}
		\label{fig:bestLinf}
	\end{figure}

\begin{figure}[H]
		\centering
		\includegraphics[width=.5\linewidth]{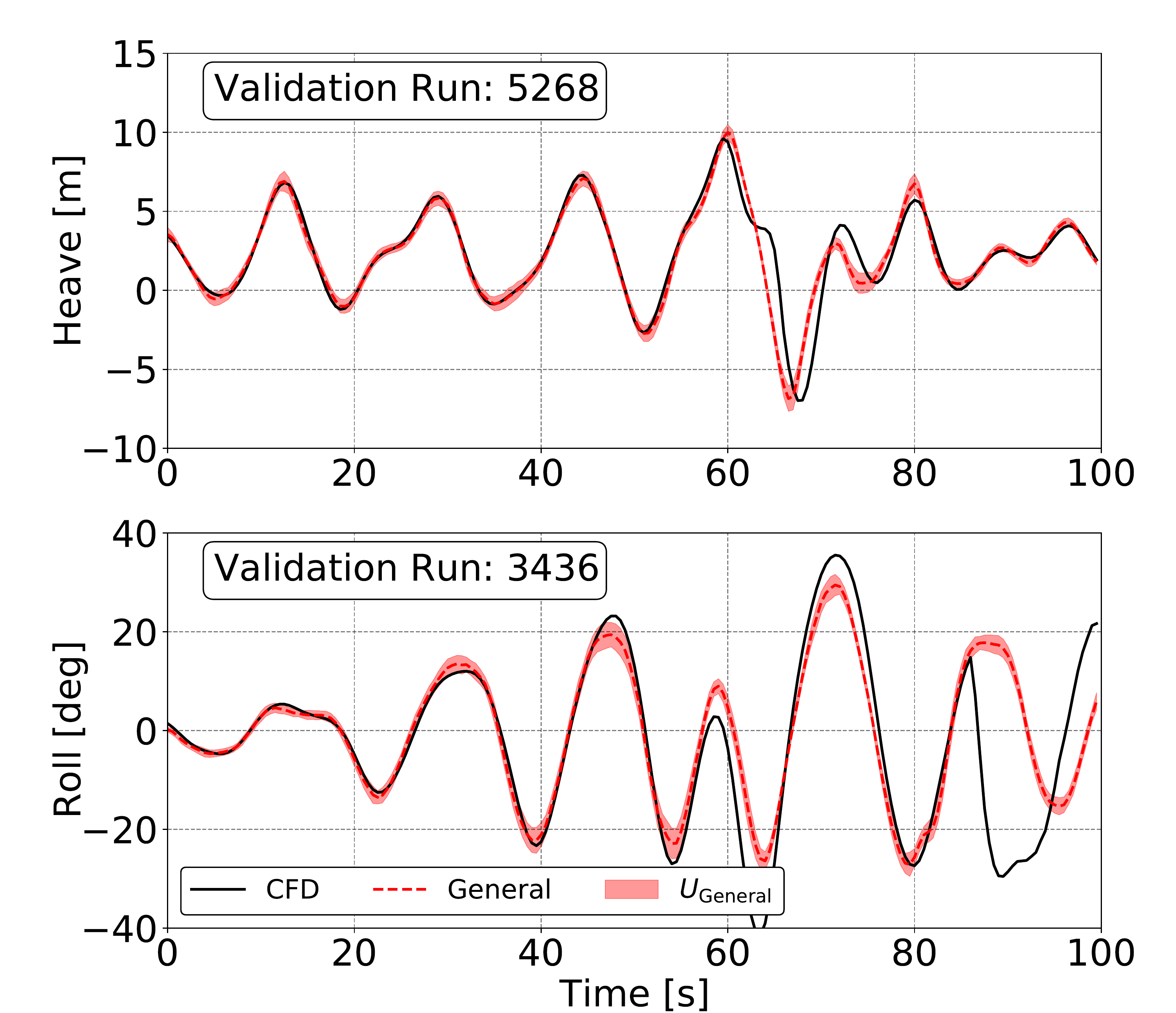}  
		\caption{Comparison of time-histories with the largest $L_\infty$ error for heave and roll with a model trained with 400 simulations.}
		\label{fig:worstLinf}
	\end{figure}

	Figs.~\ref{fig:L2} and \ref{fig:Linf} showcase the overall performance and accuracy of the neural network in terms of $L_2$ and $L_\infty$ but do not show the actual temporal LSTM prediction error. Fig.~\ref{fig:bestLinf} demonstrates the validation runs that resulted in the smallest $L_\infty$ error for heave and roll for the general approach model trained with 400 runs. CFD is compared in Fig.~\ref{fig:bestLinf} to the LSTM predictions with uncertainty estimates from the Monte Carlo Dropout approach that corresponds to two standard deviations. The LSTM predictions are able to match both the phasing and magnitude of the CFD predictions well. Fig.~\ref{fig:worstLinf} shows the validation runs with the largest $L_\infty$ error. For both heave and roll, portions of the LSTM predictions match the CFD well, while other parts of the time-history are not as well predicted, especially after significant response magnitudes. Overall, the uncertainty is on the order of 1-2~deg for the largest roll angles and is typically larger at the response peaks.
	
\begin{figure}[H]
		\centering
		\begin{subfigure}{0.49\textwidth}
			\centering
			\includegraphics[width=0.99\textwidth]{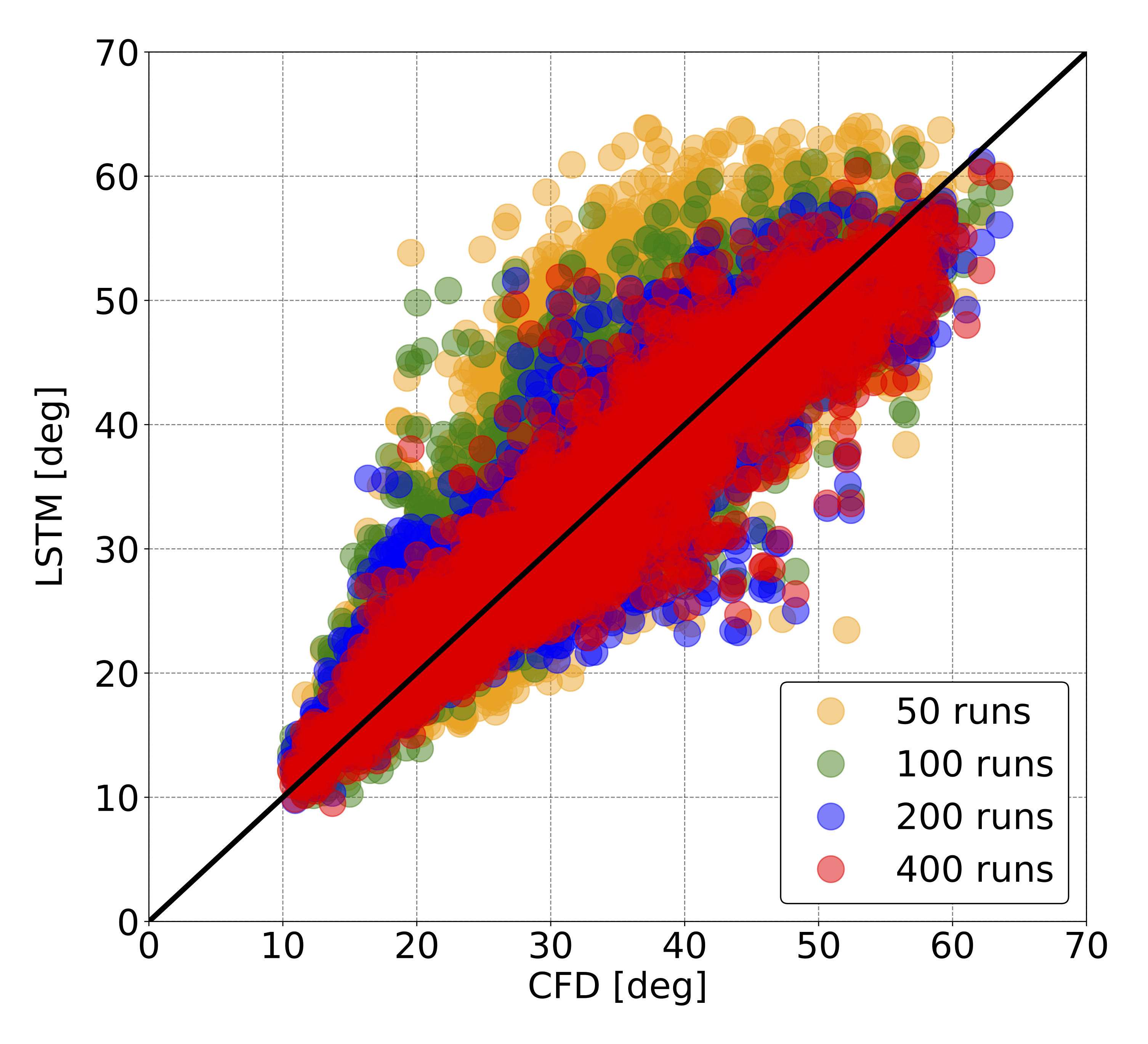}
			\caption{Ensemble}
		\end{subfigure}
		\begin{subfigure}{0.49\textwidth}
			\centering
			\includegraphics[width=0.99\textwidth]{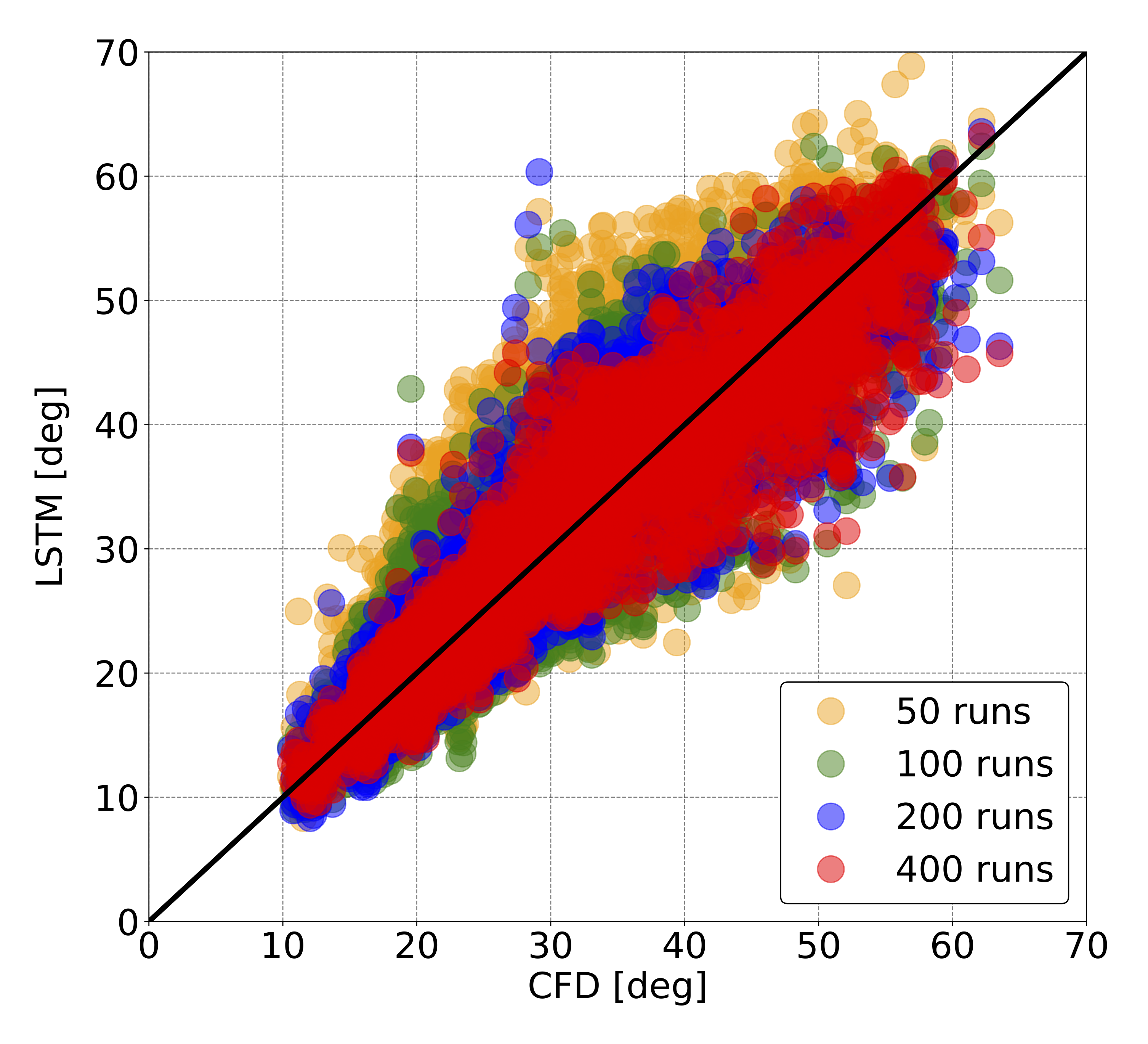}
			\caption{General}
		\end{subfigure}
		\caption{Comparison of the absolute maximum roll for each composite wave run with CFD and LSTM models with varying amounts of training data}
		\label{fig:compareMax}
	\end{figure}

	Fig.~\ref{fig:L2} through \ref{fig:worstLinf} provide overall assessments on the accuracy of the LSTM models to reproduce the temporal response of the CFD simulations. However, the CWG methodology is concerned with the extremes, and therefore, the absolute maximums are of greater importance than the temporal predictions. Although, the prediction time-histories can provide insight into the mechanisms causing extremes. The CFD and LSTM predictions of the maximum roll due to each composite wave train are compared in Fig.~\ref{fig:compareMax} for the ensemble and general approaches. Each marker in Fig.~\ref{fig:compareMax} for a particular model corresponds to a single composite wave train and the corresponding CFD and LSTM predictions. The solid black line denotes identical values for CFD and LSTM. Like the comparisons of $L_\infty$, the 400 training run models follows similar trends, while there seems to be more spread in the data for models trained with less data. Both approaches demonstrate convergence towards better correlation.
	
\begin{figure}[H]
		\centering
		\begin{subfigure}{0.49\textwidth}
			\centering
			\includegraphics[width=0.99\textwidth]{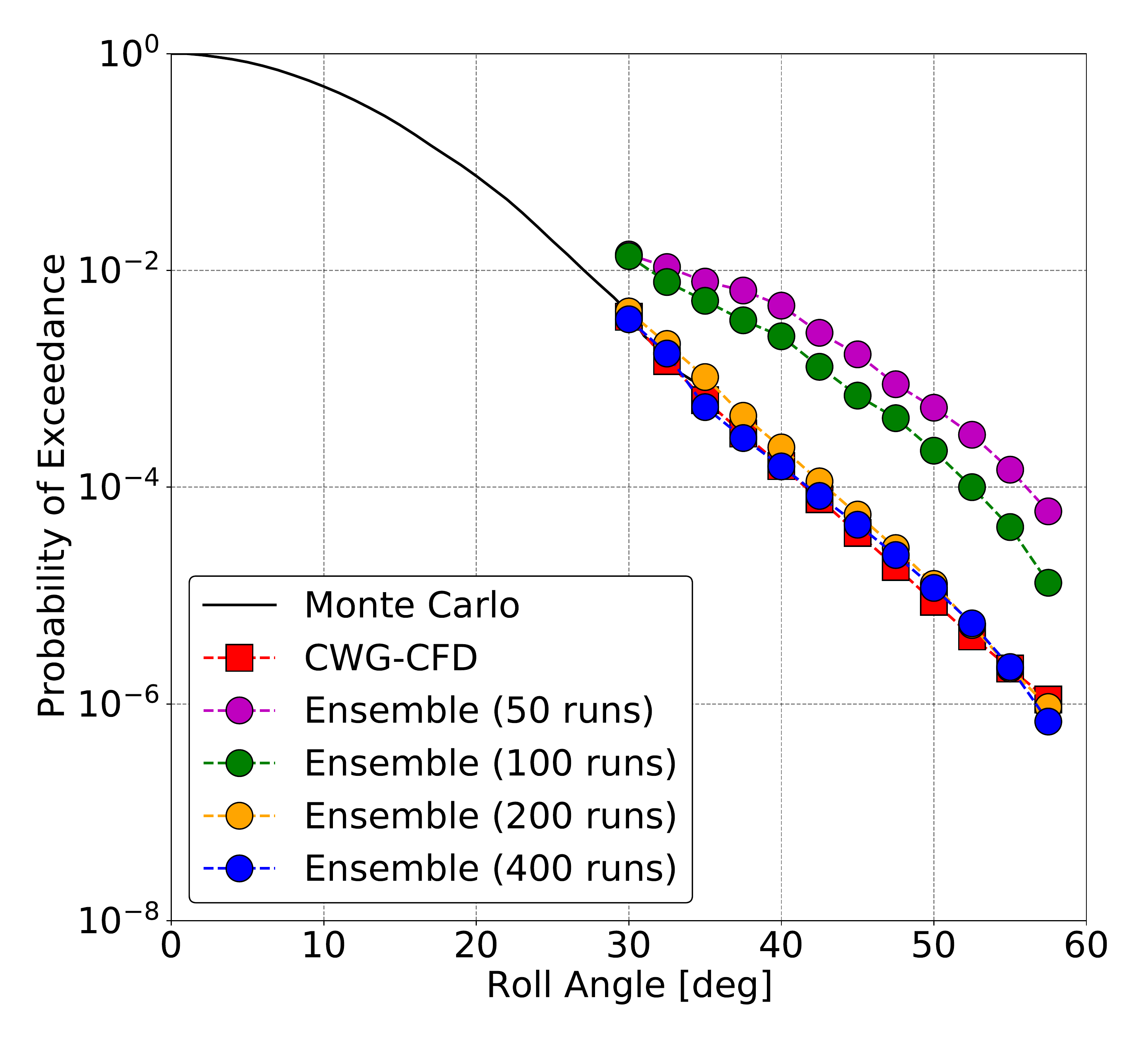}
			\caption{Ensemble}
		\end{subfigure}
		\begin{subfigure}{0.49\textwidth}
			\centering
			\includegraphics[width=0.99\textwidth]{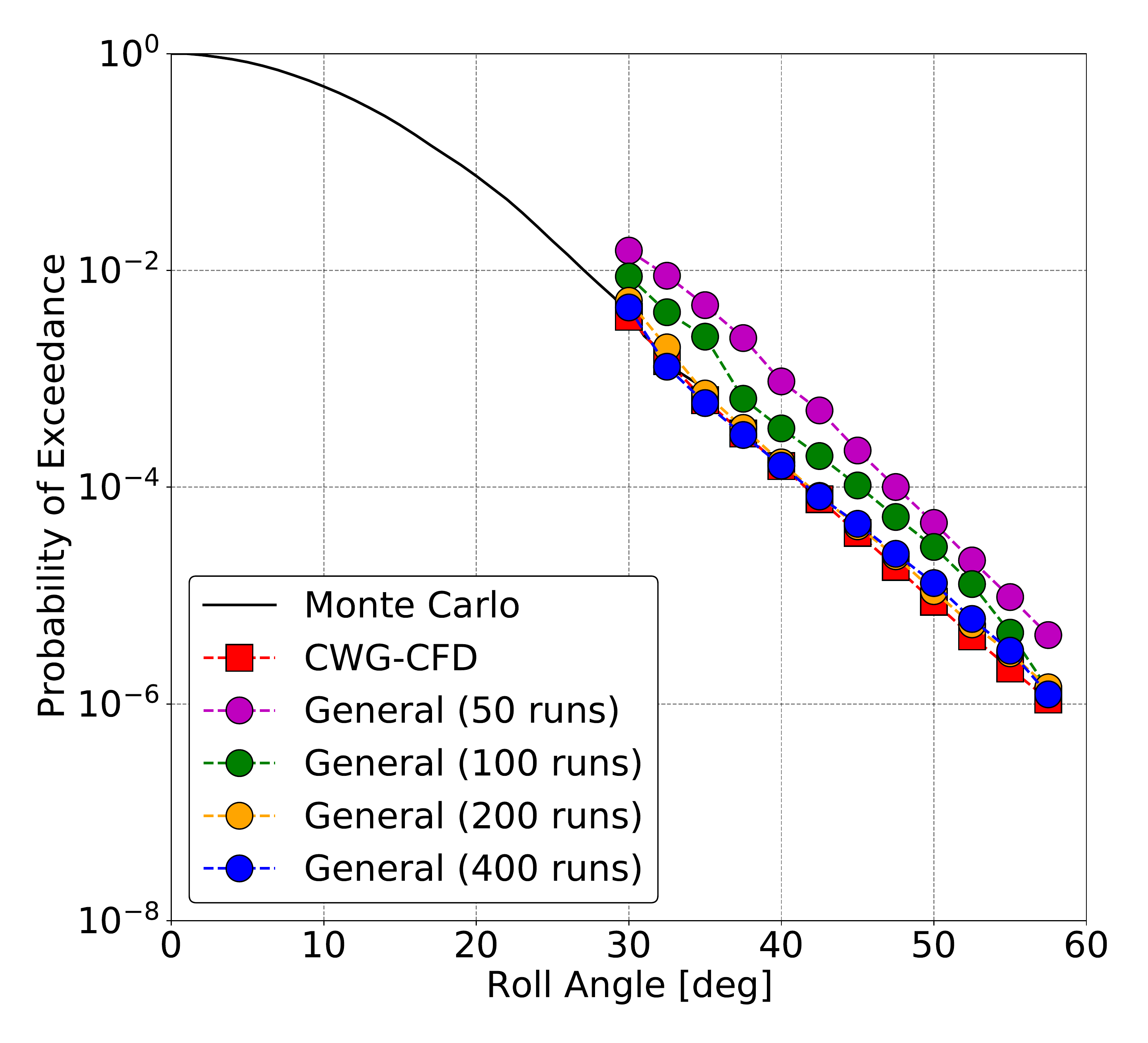}
			\caption{General}
		\end{subfigure}
		\caption{Probability of exceedance of roll in Sea State 7.}
		\label{fig:probExceed}
	\end{figure}

	Fig.~\ref{fig:compareMax} shows how the absolute maximum roll compares for each individual composite wave train. The probability of occurrence of the composite wave trains calculates the probability of exceedance shown in Fig.~\ref{fig:probExceed} for both modeling approaches. With 200 training runs, both approaches are able to represent the CWG-CFD results from \cite{Silva2021oe}. When less training data is available, the general model is more accurate. However, with at least 200~training runs available, both approaches are able to reproduce the CWG-CFD prediction of probability of exceedance.
	
\begin{figure}[H]
		\centering
		\includegraphics[width=.5\linewidth]{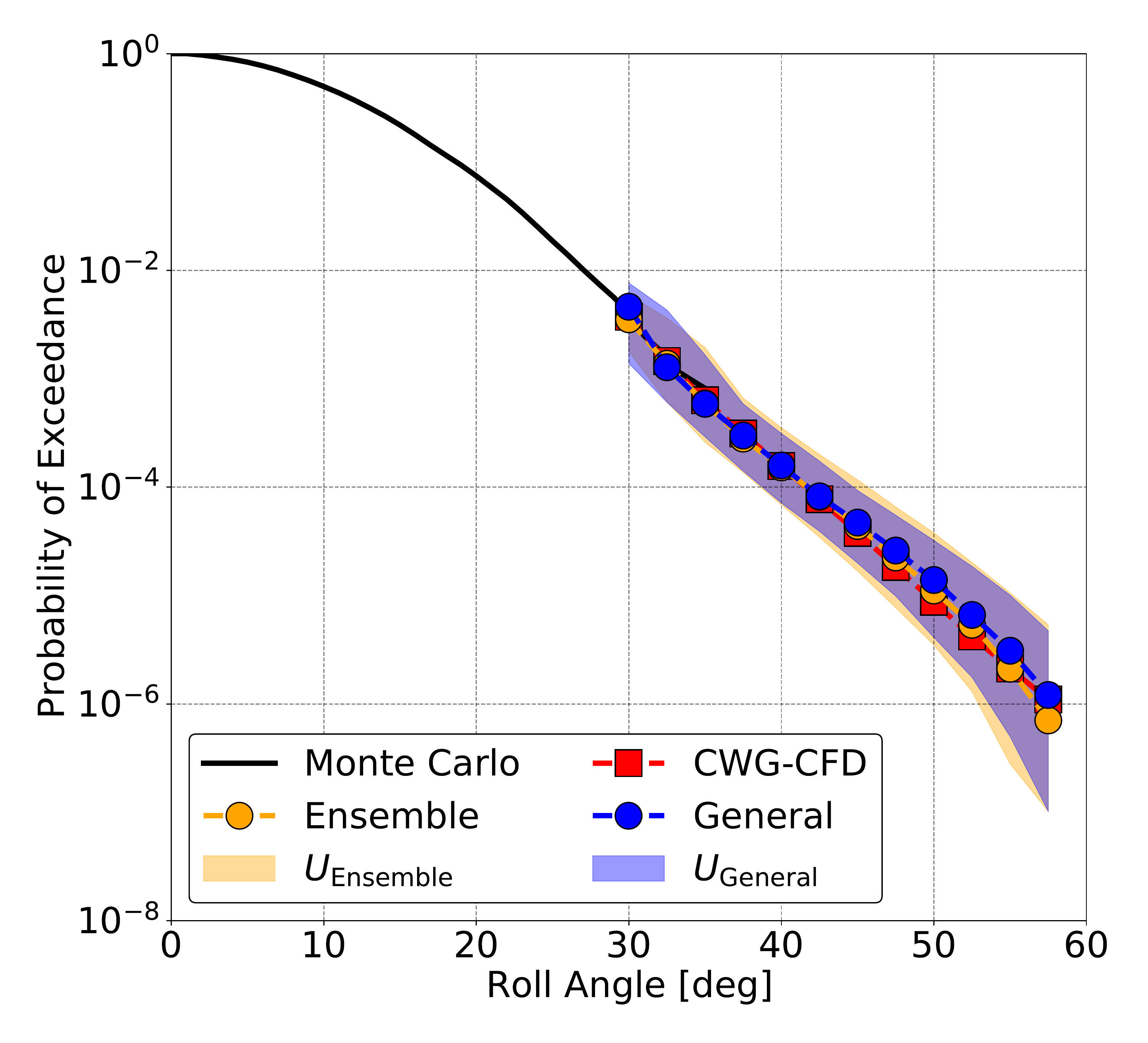}  
		\caption{Comparison of probability of exceedance with uncertainty estimates for approaches that trained with 400~runs.}
		\label{fig:probExceed_wUnc}
	\end{figure}

	Fig.~\ref{fig:probExceed_wUnc} compares the uncertainty in the LSTM predictions with both approaches for 400 total training runs. The uncertainty for both approaches increases as the roll angle of interest increases. Although the uncertainty is low in the time-history predictions in Fig.~\ref{fig:L2} through \ref{fig:worstLinf}, the uncertainty compounds across all the considered composite wave trains. The compounding of uncertainty and the sensitivity of probability of exceedance calculation yields uncertainty in the probability calculation that is of the same order of magnitude as the probability itself at the largest roll angles. The large uncertainty highlights that as a response of interest becomes more extreme and more rare, the probability calculation is detrimentally sensitive to underlying error.

	The CWG-CFD method in \cite{Silva2021oe} reduces the computational cost compared to the Monte-Carlo simulation for the same case study by roughly five~orders of magnitude at a roll angle of 57.5~deg (Fig.~\ref{fig:probExceed}). This reduction in computational cost is with respect to the estimated exposure time needed to observe a given most probable maximum roll angle with Eqn.~\eqref{eq:mpm-ochi} from \cite{Ochi1998}. In Eqn.~\eqref{eq:mpm-ochi}, $\overline{y_n}$ is the most probable maximum response, $T$ is the exposure time in hours, $m_0$ and $m_1$ are the zeroth and second spectral moments, respectively.
	
	\begin{equation}
	\overline{y_n} \ = \ 
	\sqrt{m_0} \left[ 
	2\ln \left( \frac{ 60^2 T}{2\pi} \sqrt{\frac{m_2}{m_0}}
	\right)\right]^{\frac{1}{2}}
	\label{eq:mpm-ochi}
	\end{equation}
	
\begin{figure}[H]
		\centering
		\begin{subfigure}{0.5\textwidth}
			\centering
			\includegraphics[width=0.99\textwidth]{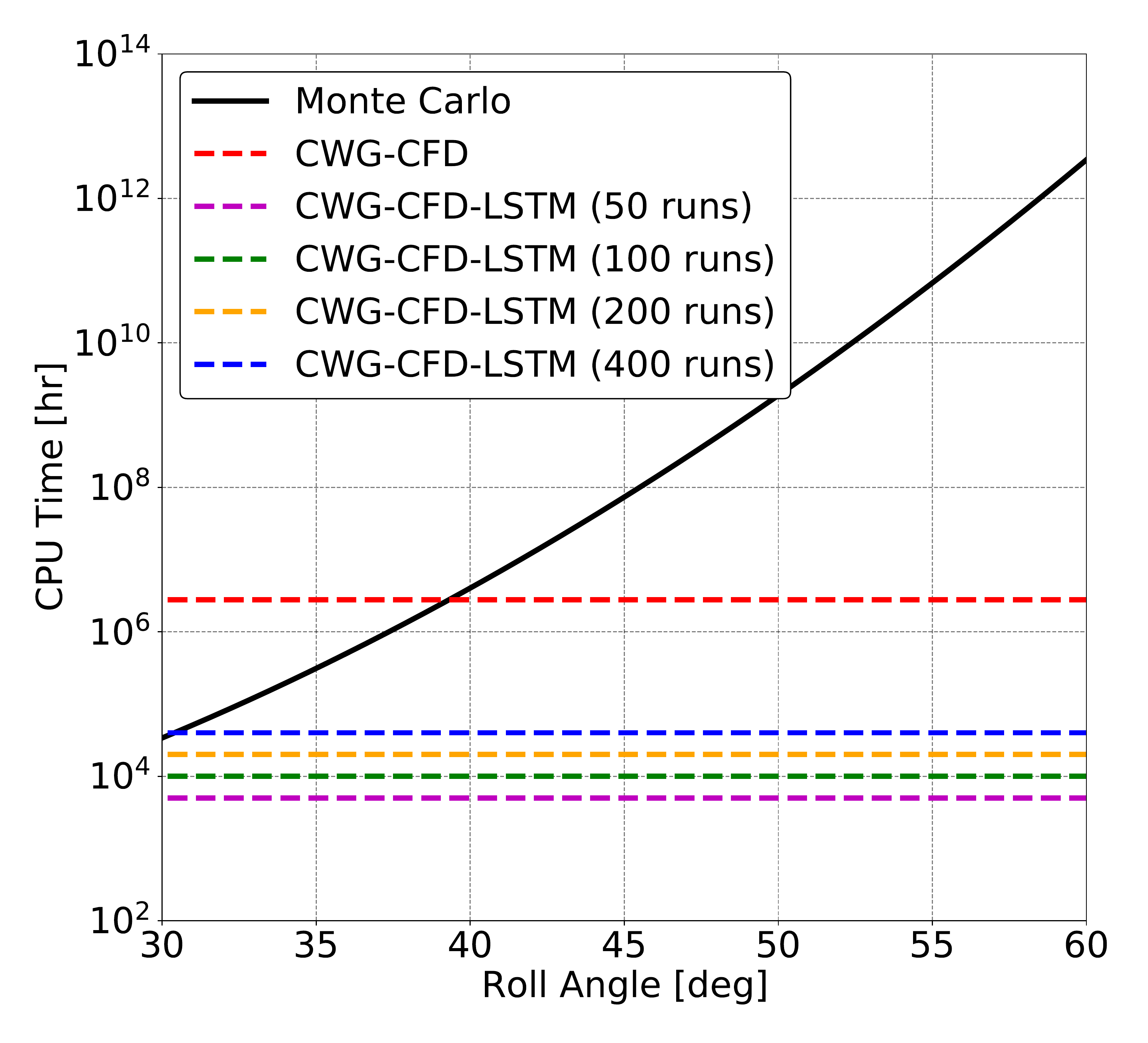}
			\caption{CPU Time}
		\end{subfigure}
		\begin{subfigure}{0.5\textwidth}
			\centering
			\includegraphics[width=0.99\textwidth]{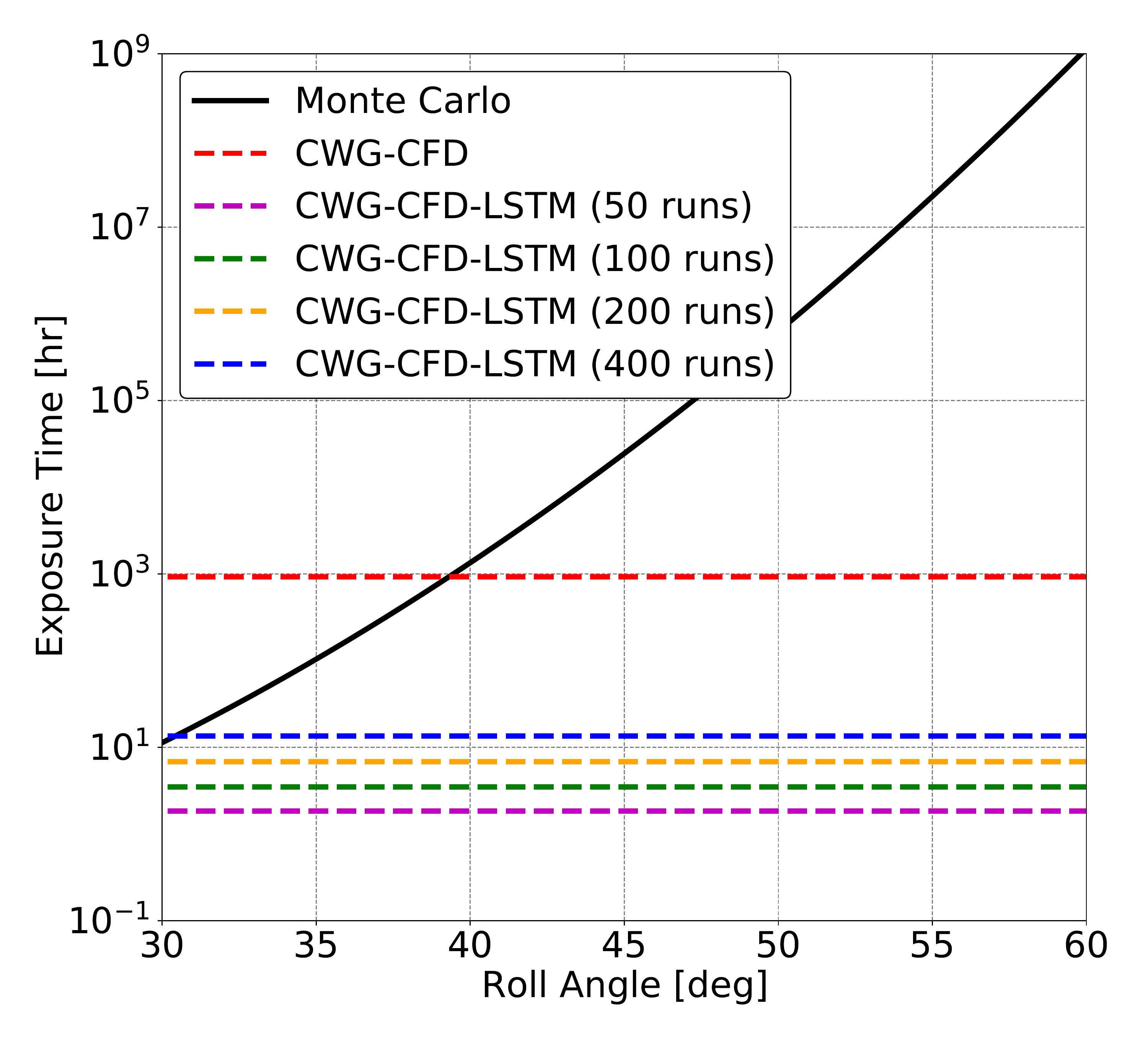}
			\caption{Exposure Time}
		\end{subfigure}
		\caption{Required CPU and exposure time for the CFD, CWG-CFD, and CWG-CFD-LSTM methods.}
		\label{fig:cost}
	\end{figure}

	The total central processing unit (CPU) and exposure time of the CWG-CFD and CWG-CFD-LSTM methods are in Fig.~\ref{fig:cost} and compared to the required simulation time for Monte Carlo analysis estimated with Eqn.~\eqref{eq:mpm-ochi}. The CPU time corresponds to the total computational cost of the methodologies and is specific to the considered mesh, software, and computing system. Meanwhile, the exposure time is the actual time simulated, which is consistent across systems and would be applicable to model testing as well as a 3-D simulation. The CPU and exposure time required of the CWG implementations also includes the random wave simulations considered in the identification of the natural initial conditions and the associated probability distribution of the encounter conditions. The CWG-CFD- LSTM model with 400 training runs, results in  two orders of magnitude reduction in the total computational cost of the CWG-CFD method. Thus, the CWG-CFD-LSTM methodology yields a total estimated reduction of seven orders of magnitude in computational cost to produce a probability of exceedance of up to a threshold of 57.5~deg. The cost of the CWG-CFD-LSTM methods is close to that of the Monte Carlo for producing probability of exceedance predictions only up to roll angles of 30~deg.
	
\begin{figure}[H]
		\centering
		\includegraphics[width=.5\linewidth]{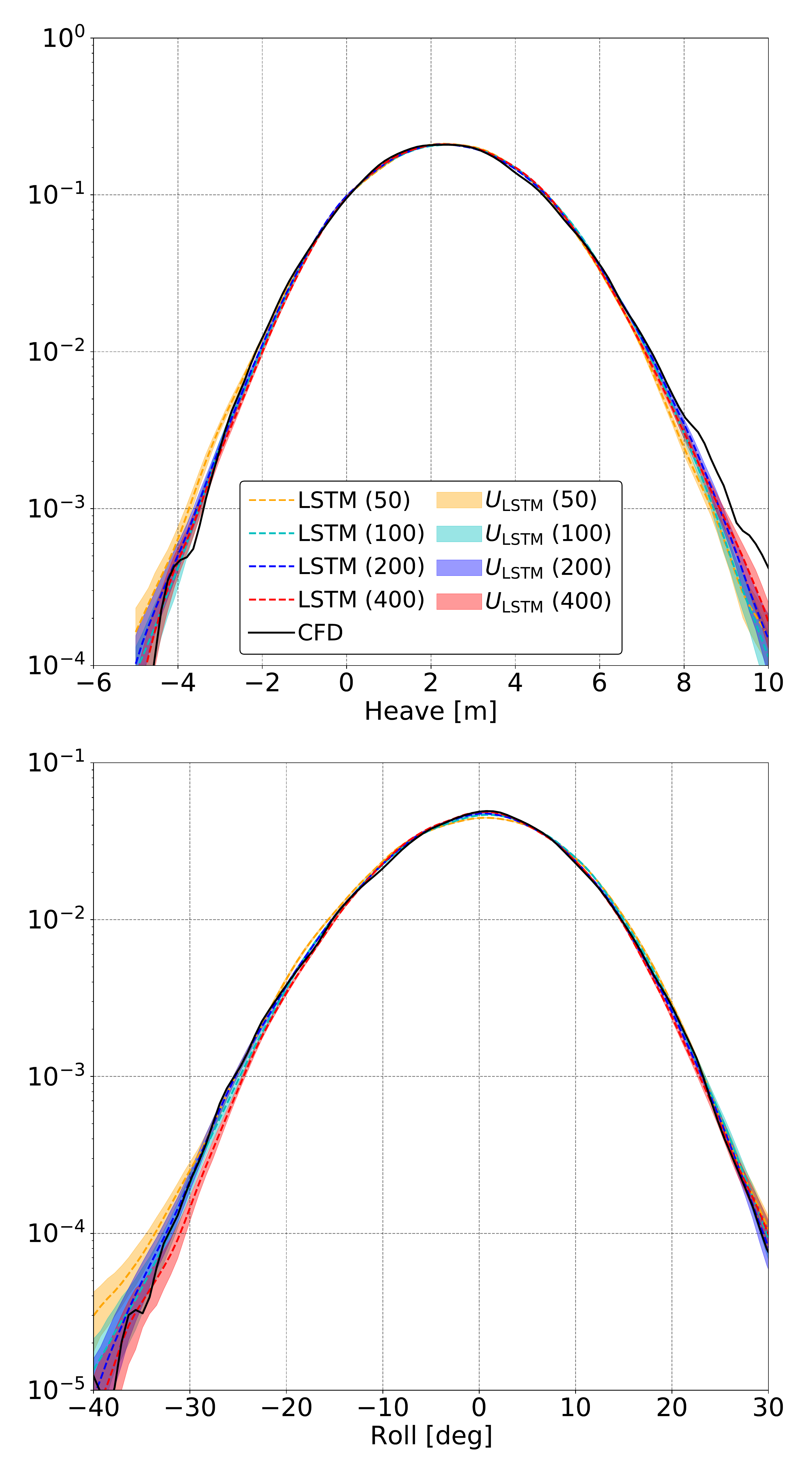}  
		\caption{Comparison of the PDF in a logarithmic scale for each DoF with models trained with the general approach.}
		\label{fig:compareDistlog}
	\end{figure}

	The current paper constitutes significant progress, not only in the development of a computationally efficient framework for extreme ship response quantification, but also demonstrates a way of training neural networks to produce real-time observations of extremes. Previous research training neural networks to represent the dynamical response of vessels has focused on random waves and has not explored the application to extremes. Neural networks in general are better at interpolation than extrapolation and perform poorly in predicting extreme events when trained with random data. However, the present neural nets are trained with composite wave trains containing large and rare deterministic wave groups that lead to extremes. Therefore, the composite wave trains can, in principle, be applied to training robust neural network surrogate models that are capable of performing Monte Carlo analysis that recovers the entire PDF of each DoF, rather than simplify the probability of exceedance that is calculated within the CWG method. Fig.~\ref{fig:compareDistlog} compares CFD and LSTM predictions of heave and roll PDF in logarithmic scale with only the general modeling approach, for 100 hours of exposure time. The roll and negative portion of the heave are well represented, but the positive heave is under-predicted. Similar to the probability of exceedance comparisons, the PDF starts to converge around 200~training runs. Although the LSTM model performs well and is able to recover the underlying PDF, the presented CFD results only consider a 100~hour exposure window, which is not enough to induce significant extremes. The exposure time projections predicted with Eqn.~\eqref{eq:mpm-ochi} and shown in Fig.~\ref{fig:cost} predicts a most probable maximum of 35~deg with a 100~hour exposure window, which qualitatively agrees with the absolute maximum roll angles observed in Fig.~\ref{fig:compareDistlog}. To achieve the larger roll angles in Fig.~\ref{fig:probExceed}, significantly more exposure time is required. Therefore, although the prediction of the PDF with the LSTM neural networks is encouraging, further research is needed to evaluate whether CWG-CFD-LSTM training methodology is suitable for recovering the entire extreme PDF through a Monte Carlo prediction with the neural network.
		
	\section*{Conclusion}
	
	A framework is presented that expands upon the research of \cite{Silva2021oe} and \cite{Xu2021} to predict extreme response statistics with the CWG method, CFD, and LSTM neural networks. The new CWG-CFD-LSTM framework is demonstrated with a case study of a 2-D midship section of the ONRT in Sea State 7. The results of the framework are compared to a CWG-CFD method with various amounts of training data, with a general approach, where a single neural network model is trained for all composite wave trains as well as an ensemble approach, where multiple models are trained, each responsible for composite wave trains that contain a wave group with the same $T_c$ and $j$. Both approaches are able produce motions responses and probabilistic predictions that are representative of the CWG-CFD method, with 200~total training runs, but with two orders of magnitude of computational cost savings. The CWG-CFD-LSTM framework in total, results in an estimated reduction of seven orders of magnitude in computational cost compared to a Monte Carlo type approach. The work is an important step forward in developing a generalized framework that renders the CWG method accessible to both CFD and experiments with LSTM as a surrogate to represent the underlying dynamical processes.
	
	\section*{Acknowledgments}
	
	This work is supported by the Department of Defense (DoD) Science, Mathematics, and Research for Transformation (SMART) scholarship, the Naval Surface Warfare Center Carderock Division (NSWCCD) Extended Term Training (ETT), and the NSWCCD Naval Innovative Science and Engineering (NISE) programs. The authors would also like to acknowledge and thank the Office of Naval Research for its support of this research under contracts N00014-20-1-2096, led by the program manager Woei-Min Lin.

	\bibliographystyle{abbrvnat}
	\bibliography{References}  
	
\end{document}